\documentclass[twocolumn,english,preprintnumbers,amsmath,amssymb,pre]{revtex4}

\usepackage{graphicx}
\usepackage{color}

\begin{document}

\title{Isotopic effects in chair graphane}
\author{Carlos P. Herrero}
\author{Rafael Ram\'irez}
\affiliation{Instituto de Ciencia de Materiales de Madrid,
         Consejo Superior de Investigaciones Cient\'ificas (CSIC),
         Campus de Cantoblanco, 28049 Madrid, Spain }
\date{\today}

\begin{abstract}
Graphane is a layered material consisting of a sheet 
of hydrogenated graphene, with a C:H ratio of 1:1.
We study isotopic effects in the properties of chair 
graphane, where H atoms alternate in a chairlike arrangement
on both sides of the carbon layer. 
We use path-integral molecular dynamics simulations,
which allows one to analyze the influence of nuclear
quantum effects on equilibrium variables of materials.
Finite-temperature properties of graphane are studied in 
the range 50--1500~K as functions of the isotopic mass of the 
constituent atoms, using an efficient tight-bonding potential. 
Results are presented for kinetic and internal energy, 
atomic mean-square displacements, fluctuations in the C--H 
bond direction, plus interatomic distances and layer area.
At low temperature, substituting $^{13}$C for $^{12}$C gives
a fractional change of $-2.6 \times 10^{-4}$ in C--C distance 
and $-3.9 \times 10^{-4}$ in the graphane layer area.
Replacing $^2$H for $^1$H causes a larger fractional 
change in the C--H bond of $-5.7 \times 10^{-3}$.
The isotopic effect in C--C bond distance increases 
(decreases) by applying a tensile (compressive) in-plane stress.
These results are interpreted in terms of a quasiharmonic 
approximation for the vibrational modes.
Similarities and differences with isotopic effects in graphene 
are discussed.   \\

\end{abstract}

\maketitle

\section{Introduction}

Graphane is a quasi-two-dimensional structure of carbon atoms 
arranged in a buckled honeycomb lattice covalently bonded to 
hydrogen atoms.
There exist several conformers of graphane, the most studied
of them being the so-called chair graphane. In this structure,
hydrogen atoms alternate in a chairlike pattern on both
sides of the carbon layer, with a stoichiometric
C:H ratio equal to 1 \cite{so07,jo10,we11}.
In Fig.~1 we show a ball-and-stick model of chair graphane,
displaying C and H atoms as large yellow and small blue balls, 
respectively. 

\begin{figure}
\vspace{-0.6cm}
\includegraphics[width=7cm]{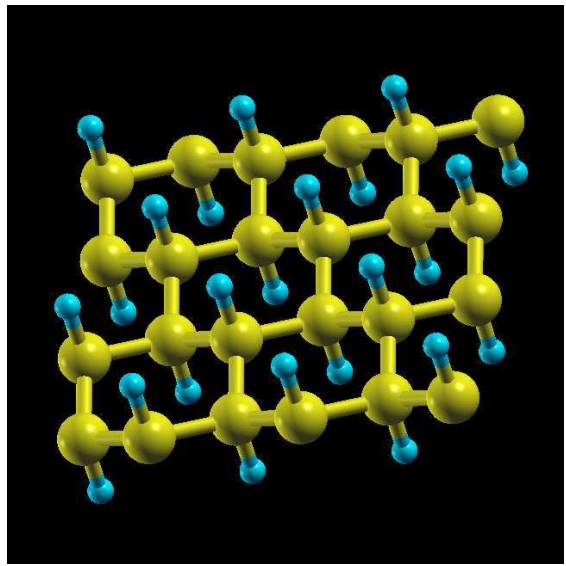}
\vspace{-0.6cm}
\vspace{1cm}
\caption{Ball-and-stick image of chair graphane.
Large yellow and small blue balls represent carbon and
hydrogen atoms, respectively.
}
\label{f1}
\end{figure}

Several forms of hydrogenated graphene, and chair graphane in 
particular, have been studied in detail in the last few years.
It is known that graphane may be formed by chemisorption
of hydrogen on graphene in a reversible way \cite{el09}.
This process gives rise to a rearrangement of the
interatomic bonds and angles in the graphene structure.
Each H atom bounds to a C atom, so the latter changes
its sp$^2$ orbital hybridization to sp$^3$,
and therefore the planar structure of graphene transforms
into an out-of-plane buckled configuration.
Graphane turns out to be a wide band-gap semiconductor, 
and an important spin polarization may be obtained by 
creating domains of H vacancies or CH divacancies \cite{sa10b}.
Metal dopants or alkaline atoms have been shown to change its
magnetic and electronic properties \cite{en13,wa16b,ma17b}.

Many properties of crystalline solids can be calculated with 
reasonable good accuracy by using some kind of harmonic
approximation for the vibrational modes in the material.
However, anharmonic effects are crucial to describe important
properties, such as thermal expansion and pressure 
dependence of the compressibility.
Another important consequence of anharmonicity is the
isotope dependence of structural properties of materials and 
of their melting temperature \cite{as76,ki96,ra10}.

The influence of isotopic composition on structural parameters, 
lattice dynamics, and electronic properties of three-dimensional
materials has been analyzed along the years by several experimental 
and theoretical techniques \cite{ca93,ca00,he09c,he11}.
Isotopic effects in structural properties of solids,
apart from pure scientific interest, are relevant in
various fields such as metrology. In this field, high-accuracy
measurements of lattice constants in crystals with
isotopically-controlled composition are important for
a precise determination of the kilogram \cite{an11b} and
the Avogadro constant \cite{an11,az15}.

In recent years, various isotopic effects were studied
in 2D materials, mainly graphene \cite{be12,co13,br14,hu10,da17}.
For graphane, a controlled hydrogen isotope composition
can be achieved \cite{ba13}. In this case, isotopic effects have
been studied for vibrational and thermodynamic properties, mainly
using density-functional perturbation theory \cite{hu13b,hu14}.
Moreover, the influence of the H isotopic mass on the kinetics of the
graphene hydrogenation process has been studied in detail \cite{pa13,wa17}.

Several types of isotopic effects can be studied in crystalline 2D and
3D materials, among which particular relevance is associated to
those due to the change of phonon frequencies with the
atomic mass \cite{ca00}. 
The mass dependence of the frequencies causes changes in the 
vibrational amplitudes. At low $T$, these amplitudes are larger
for smaller atomic mass (quantum zero-point motion), but at high $T$
they become independent of the mass (classical limit). 
For larger vibrational amplitudes, the atomic motion {\em explores} more
effectively the anharmonicity of the interatomic potential, which
gives a mass-dependence for various thermodynamic and structural
properties.

Isotopic effects in equilibrium properties of crystalline solids
are due to a combination of both, the quantum nature of atomic nuclei 
and the anharmonicity of the interatomic potentials.
Then a convenient theoretical framework to study this question is 
the Feynman path-integral formulation for the statistical mechanics 
of many-body quantum systems at finite temperature \cite{fe72,kl90,ce95}.
In this line, the path-integral molecular dynamics (PIMD) method is
a powerful technique to study problems where anharmonic and 
quantum effects are relevant.
This kind of atomistic simulations have been employed 
earlier to study isotopic effects in solids, in particular 
for structural parameters and thermodynamic properties
\cite{mu95,he99,he09c}.

In this paper, we study isotopic effects in chair graphane 
by PIMD simulations in a temperature range from 50 to 1500~K, 
using an efficient tight-binding (TB) Hamiltonian,
which has been found to accurately describe various structural
and thermodynamic properties of carbon-based 
materials \cite{lo09,ra16,he16}.
We consider the isotopes $^{12}$C and $^{13}$C for carbon,
along with $^1$H and $^2$H for hydrogen.
We quantitatively analyze the influence of isotopic mass on 
structural properties such as C--C and C--H interatomic distances,
as well as on the area of the graphane layer, which displays 
a temperature dependence very different from the classical 
approximation. The quantum motion is analyzed by studying
the atomic mean-square displacements.
The isotope effect on the C--C distance is also found to appreciably
change in the presence of an applied stress (tensile or compressive).
Graphane made up with the most abundant isotopes, $^{12}$C and $^1$H,
is taken as a reference for the isotopic effects in the variables
studied in this paper.
Results derived from our PIMD simulations are compared with
calculations based on a quasiharmonic approximation
(QHA) for the vibrational modes.

 The paper is organized as follows. In Sec.\,II we describe the
computational techniques employed here, i.e., PIMD method and 
tight-binding procedure.
In Sec.\,III we present results for the internal energy of graphane,
with particular emphasis on the kinetic energy.
Isotopic effects in the interatomic distances for C--C and C--H bonds
are discussed in Sec.~\,IV.
In Sec.~\,V we analyze the atomic mean-square displacements and
quantum delocalization. Isotopic effects in the in-plane area 
of the graphane sheet are presented in Sec.\,VI, and the effect of
an external stress is discussed in Sec.\,VII.
The paper closes in Sec.\,VIII with a summary.

\section{Method}

In this paper we study the dependence of equilibrium properties of
chair graphane on isotopic mass. This dependence is not present in
classical calculations, irrespective of the anharmonicity
in the atomic vibrations.
This is well known for 3D solids, and can be directly
derived from basic arguments of classical statistical mechanics.
In fact, in this case equilibrium properties that depend only on the
coordinates (positions) of the particles do not change with the
mass $M$ \cite{re65,as76,he20c}.
In quantum statistical physics, momenta and positions do not commute
and, as a consequence, the atomic mass $M$ affects the mean values of
position-dependent variables.
Thus, isotopic effects in equilibrium properties are due to
the quantum dynamics of atomic nuclei.
The same reasons apply for 2D materials, as graphane studied here.
In addition, they are anharmonic effects, like thermal expansion, 
as they do not show up in the absence of anharmonicity in 
the interatomic potentials.

We employ the PIMD method to study equilibrium 
properties of chair graphane at various temperatures.
$^{12}$C and $^{13}$C isotopes are considered for carbon, 
as well as $^1$H and $^2$H for hydrogen. 
The PIMD technique is based on the fact that the partition function
of a quantum system can be written in a fashion similar to that
of a classical system, which is realized by substituting each
quantum particle of mass $M$ by a ring polymer made up of 
$N_{\rm Tr}$ classical particles ({\em beads}), connected by 
springs with constant $k_{\rm har} = M N_{\rm Tr} / \beta^2 \hbar^2$
($N_{\rm Tr}$ is the so-called {\em Trotter number} and
$\beta = 1 / (k_B T)$)  \cite{fe72,gi88,kl90,ce95}.  
Such an {\em isomorphism} between quantum and classical systems
becomes formally exact in the limit $N_{\rm Tr} \to \infty$.
Details of this simulation technique and applications to
condensed-matter systems can be found
elsewhere \cite{ch81,gi88,tu10,he14}.

We use PIMD simulations to sample the phase space 
associated to the classical isomorph of our quantum system
($N$ pairs C--H).
In this procedure, the atomic dynamics does not correspond
to the quantum dynamics of the actual particles.
It is, however, effective to accurately sample the
configuration space, and gives precise values for
time-independent properties of the considered quantum system.
Note that for our present purposes it is equally valid
a Monte Carlo sampling of Feynman path 
integrals \cite{bi10,mu95,br15,ha18,br20}.

Our simulations were carried out in the isothermal-isobaric 
ensemble $N \tau T$, where one fixes the number of atoms 
($N$ pairs C--H), the in-plane applied stress ($\tau$), 
and the temperature ($T$).
Here, the stress $\tau$ in the $(x, y)$ plane has units of
force per unit length (i.e. N/m or eV/\AA$^2$), and corresponds
to the so-called frame or mechanical tension in 
the literature \cite{sh16,fo08}.
Most of our simulations have been carried out for unstressed
graphane, i.e. $\tau = 0$, but some of them were performed for
compressive ($\tau > 0$) or tensile stress ($\tau < 0$) to
analyze the influence on the isotopic effects studied here
(see Sec.~VII).

Effective algorithms for performing the PIMD simulations in 
the isothermal-isobaric ensemble have been employed \cite{ma99,tu10}.
We used staging variables to define the bead coordinates, and
the constant-temperature ensemble was obtained by connecting
Nos\'e-Hoover thermostats to each staging coordinate.
An additional chain of four thermostats was coupled to the barostat 
to yield the equilibrium fluctuations of the in-plane area of 
the simulation cell at the required stress $\tau$ \cite{tu10,he14}. 
The equations of motion were integrated by employing the reversible
reference system propagator algorithm (RESPA), where one can
define different time steps for the integration of slow and fast
degrees of freedom \cite{ma96}.
The equations of motion used in our simulations are
given in detail elsewhere \cite{ra20}.
We have used a time step $\Delta t$ = 0.5 fs for the dynamics
associated to the interatomic forces, which has been found 
to describe correctly the atomic motion in this problem, 
in particular the C--H stretching frequency of 
about 3000~cm$^{-1}$ \cite{ra20}.
For the dynamical equations associated to bead interactions and
thermostats, we employed a time step $\delta t = \Delta t/4$,
as in earlier simulations \cite{he06,he07}.
To check the convergence of our results with the time step, 
we have carried out some PIMD simulations with $\Delta t$ = 0.2~fs.
Differences between results obtained for $\Delta t$ = 0.2 and 0.5 fs
were much smaller than statistical error bars of
the variables considered here (e.g., kinetic and potential energy,
as well as interatomic distances).

We have studied chair graphane using simulation cells with $2N$ 
atoms, at temperatures between 50 and 1500~K.
To check the convergence of the results at low-temperatures, we
carried out also some simulations at $T = 25$~K.
The results presented here will refer to $N = 96$. Some additional 
simulations for cells with $N = 216$ were performed
to check the consistency of our results.
For the sake of comparison with results of PIMD simulations,
we also carried out classical molecular dynamics (MD) 
simulations of chair graphane (this corresponds in our
context to $N_{\rm Tr}$ = 1).
In the PIMD simulations, we took $N_{\rm Tr} \, T$ = 6000~K,
i.e. the Trotter number (number of beads) scales as the inverse
temperature \cite{he06}. 
The simulation cells had rectangular shape, with similar side length
in the $x$ and $y$ directions ($L_x \approx L_y$), along which
periodic boundary conditions were assumed.
For the out-of-plane $z$-direction, free boundary conditions
were assumed, thus simulating a free-standing layer.
For each considered temperature, a simulation run consisted of
$10^5$ PIMD steps for system equilibration, and
$2 \times 10^6$ steps for calculation of ensemble average properties.

An important question in the PIMD method is a reliable description
of the interatomic interactions.
The use of self-consistent potentials based on the Hartree-Fock method
or density-functional theory needs a large computational power,
so that the accessible size of the simulation cells and/or the number of
PIMD steps would be small, and therefore the size effects and
statistical noise would be large.  In this respect, 
we take a convenient compromise between precision and
manageability by using an efficient tight-binding (TB) Hamiltonian,
as that developed from density functional calculations 
by Porezag {\em et al.} \cite{po95}, which is especially adequate
for carbon systems.
In this line, Goringe {\em et al.} \cite{go97} reviewed the potential 
of TB methods to describe various properties of solids and molecules.
For our present purposes, it is worthwhile mentioning that
the TB potential employed here \cite{po95} was checked earlier by 
comparing its predictions for the frequencies of C--H vibrations in 
hydrocarbon molecules with their actual experimental values, and 
the results were acceptable \cite{he06}.
Detailed analyses of vibrational frequencies in this kind of
molecules, as derived from this TB potential, including 
mode anharmonicities, were presented elsewhere \cite{lo03,bo01}.
In condensed matter, this TB Hamiltonian has been employed earlier
to describe C--C and C--H interactions in carbon-based materials 
as diamond \cite{he06,he07}, graphite \cite{he10b}, 
and graphene \cite{he09a},

Combining path integrals with electronic structure methods has 
particular interest for our calculations, because both atomic nuclei 
and electrons are assumed to behave as quantum mechanical particles, 
so electron-phonon and phonon-phonon interactions
are precisely considered in the simulations.
In the reciprocal-space sampling of the electronic band-structure
calculation, we have only used the $\Gamma$
point (${\bf k} = 0$), because the result of employing a larger ${\bf k}$
set is an almost constant change in the total energy, with negligible 
influence on the energy differences between different atomic
configurations.
In Ref.~\cite{he20b} we have presented the convergence of the potential 
energy of graphane as a function of the cell size $N$. 

\begin{figure}
\vspace{-0.6cm}
\includegraphics[width=7cm]{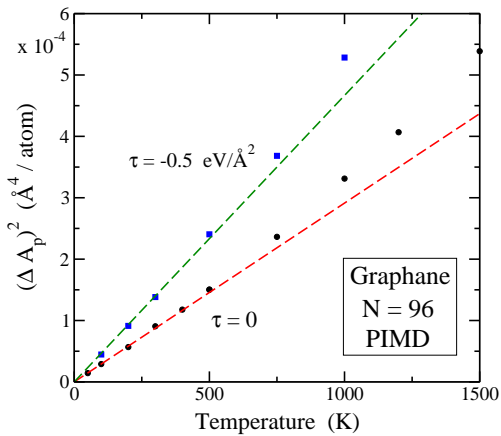}
\vspace{-0.5cm}
\caption{MSD of the layer area of chair graphane,
$(\Delta A_p)^2$, as a function of temperature.
Data points were derived from PIMD simulations for in-plane
stress $\tau$ = 0 (circles) and $-0.5$ eV/\AA$^2$ (squares).
Dashed lines indicate the linear trend displayed by
$(\Delta A_p)^2$ for $T < 500$~K.
Error bars are less than the symbol size.
}
\label{f2}
\end{figure}

In the following, we will discuss the behavior of the in-plane area
defined by the simulation cell in the $(x,y)$ plane.
We will call $A_p = L_x L_y / N$ the in-plane area per C atom.
This area fluctuates in our isothermal-isobaric ensemble.
In Fig.~2 we display the temperature dependence of the 
mean-square displacement (MSD) of $A_p$, i.e. 
$(\Delta A_p)^2 = \langle A_p^2 \rangle - \langle A_p \rangle^2$,
derived from PIMD simulations for two in-plane stresses: 
$\tau$ = 0 (circles) and --0.5 eV/\AA$^2$ (squares).
Here $\tau < 0$ indicates a tensile stress, which 
causes an increase in $(\Delta A_p)^2$ with respect to unstressed
graphane.  The dashed lines are linear fits to the data points 
obtained from simulations for $T < 500$~K.
At $T > 500$~K, the MSD of $A_p$ clearly deviates from the linear
behavior observed at lower temperatures.
This is an anharmonic effect, which grows as temperature is
raised. In fact, at $T$ = 1000~K we find an increase
of 15\% and 13\% in $(\Delta A_p)^2$ with respect to the
low-temperature extrapolation for $\tau = 0$ and --0.5 eV/\AA$^2$,
respectively.

\section{Energy}

In a classical calculation at $T = 0$ the model employed
here yields for chair graphane a layer formed by
two planar sheets of carbon atoms (corresponding to sublattices
A and B) at a distance of 0.464 \AA\ one from the other,
and two sheets of hydrogen atoms at both sides separated by
1.126 \AA\ from the nearest carbon sheet.
Then, the {\em width} of the graphane layer (separation between
H sheets) amounts to 2.716 \AA.
This configuration yields the minimum energy 
$E_0 = -57.0393$~eV/(C--H pair), which will be taken as a reference 
for our results at finite temperatures.

In a quantum calculation, the low-temperature limit shows atomic 
fluctuations due to zero-point motion, so the C and H sheets are 
not strictly planar.
In addition, anharmonicity of out-of-plane vibrations ($z$-direction)
causes a zero-point expansion, yielding a distance between
hydrogen planes of 2.739 \AA, i.e. a dilation of 0.023 \AA\ with
respect to the classical minimum.

\begin{figure}
\vspace{-0.6cm}
\includegraphics[width=7cm]{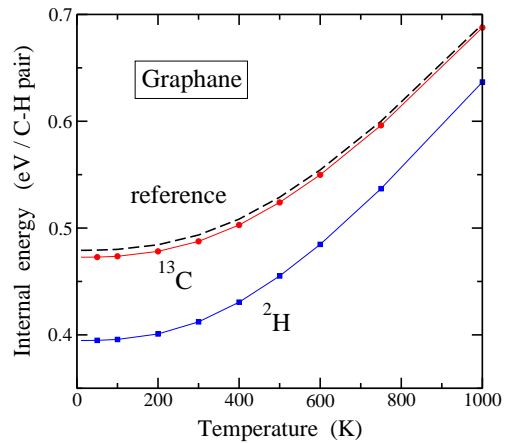}
\vspace{-0.5cm}
\caption{Internal energy of graphane vs temperature for
graphane with $^{13}$C--$^1$H (circles) and $^{12}$C-$^2$H
isotopes (squares).
The dashed line (reference) represents the result
corresponding to $^{12}$C--$^1$H graphane.
The continuous lines are guides to the eye.
}
\label{f3}
\end{figure}

In Fig.~3 we present the internal energy, $E_{\rm int}$, 
vs the temperature for chair graphane with $^{13}$C--$^1$H (circles)
and $^{12}$C--$^2$H isotopes (squares), where symbols indicate results 
of PIMD simulations.
In this kind of simulations one obtains independently 
the potential ($E_{\rm pot}$) and kinetic ($E_{\rm kin}$) energy
of the system \cite{he82,tu10,ra11}, and for $\tau = 0$
we have for the internal energy:
$E_{\rm int} = E - E_0 = E_{\rm kin} + E_{\rm pot}$.
Most of the internal energy corresponds to the vibrational
energy associated to in-plane and out-of-plane modes of graphane.
Another smaller part is the {\em elastic} energy, associated
to changes in the in-plane area $A_p$, which becomes more
appreciable at high temperature \cite{he16}.

The dashed line in Fig.~3 corresponds to the reference graphane,
made up of the most naturally abundant isotopes $^{12}$C and $^1$H.
In this case, the internal energy converges at low $T$ to
$E - E_0 =$ 479 meV/(C--H pair), which corresponds to
the zero-point energy of the material.
Replacement of $^{13}$C for $^{12}$C reduces the zero-point
energy by a small amount (7 meV per C--H pair), whereas
changing $^1$H by $^2$H causes a reduction of 85 meV,
a 18\% of the zero-point energy of the reference material.

\begin{figure}
\vspace{-0.6cm}
\includegraphics[width=7cm]{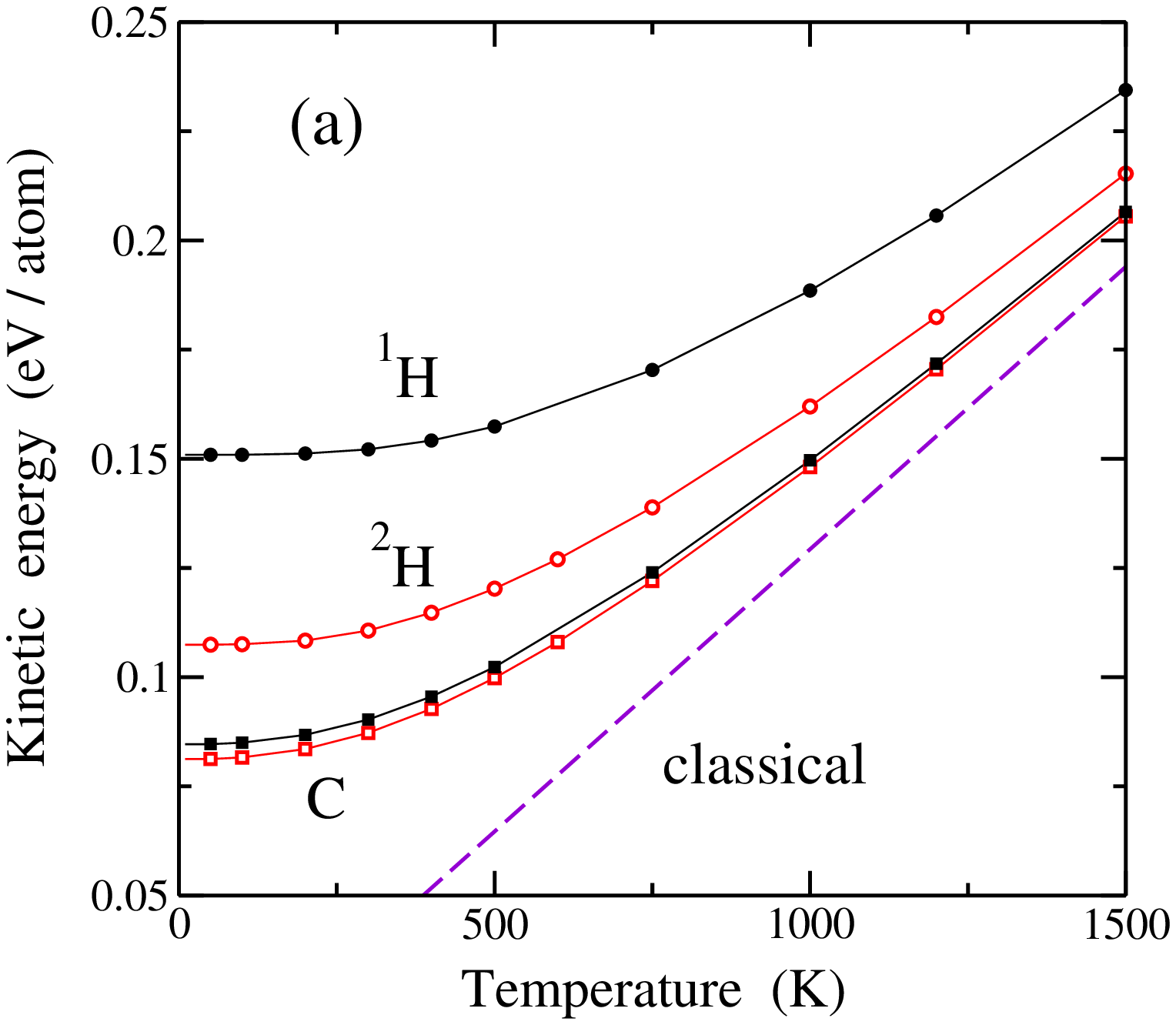}
\vspace{-0.0cm}
\includegraphics[width=7cm]{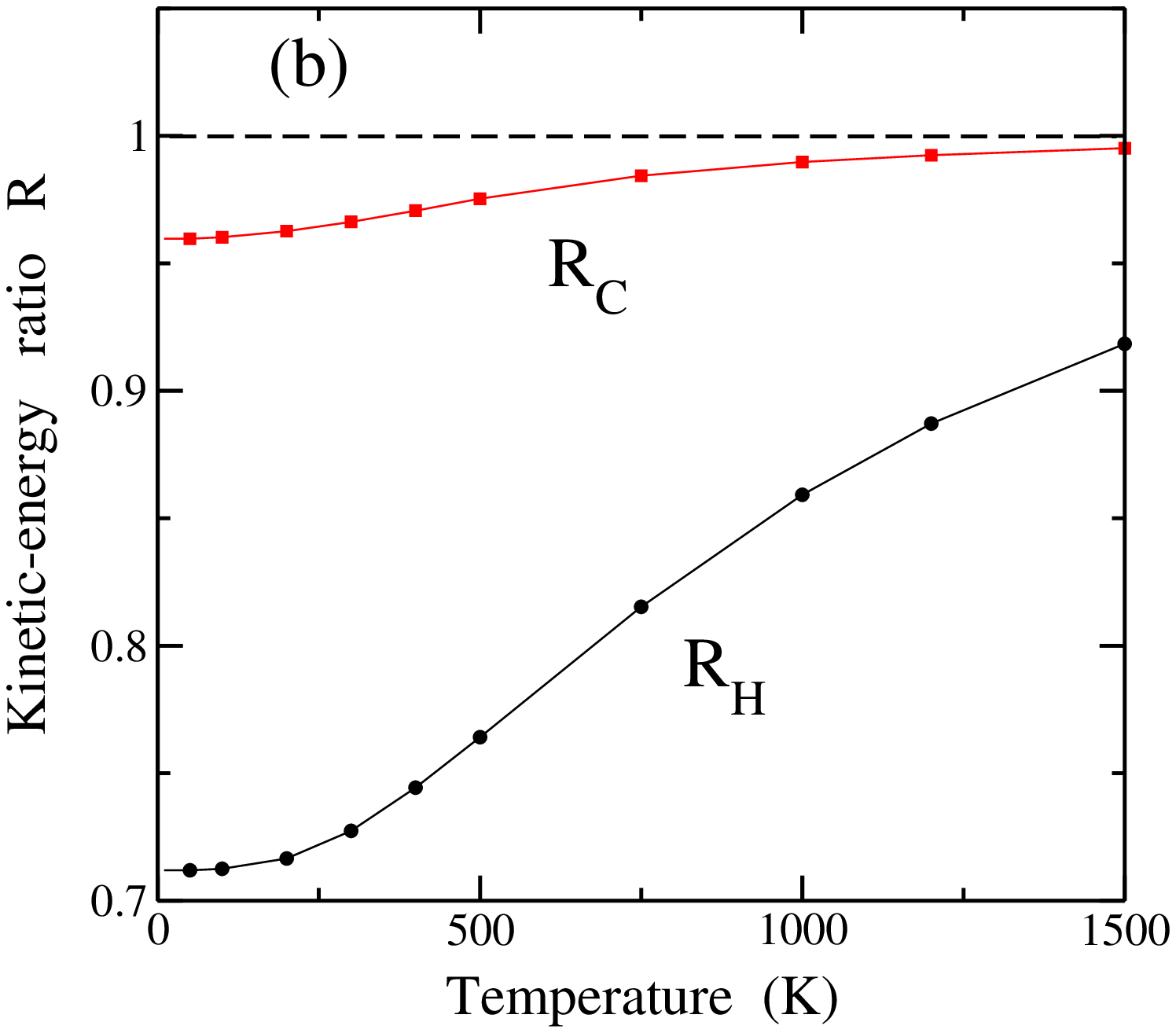}
\vspace{-0.5cm}
\caption{ (a) Temperature dependence of the kinetic energy of
hydrogen and carbon isotopes in graphane.
Symbols represent results of PIMD simulations for
$^1$H (solid circles), $^2$H (open circles),
$^{12}$C (solid squares), and $^{13}$C (open squares).
A dashed line indicates the classical limit:
$E_{\rm kin} = 3 k_B T / 2$ per atom.
(b) Temperature dependence of the ratio $R_{\rm C}$ between
kinetic energies of $^{13}$C and $^{12}$C (squares), as well as
$R_{\rm H}$ between $^2$H and $^1$H (circles), derived from
the data displayed in (a).
A horizontal dashed line shows the classical limit of
the kinetic-energy ratio.
}
\label{f4}
\end{figure}

The kinetic energy of H and C atoms is shown as a function of
temperature in Fig.~4(a). Symbols are data points derived from
PIMD simulations of chair graphane. Solid symbols correspond to 
the lighter isotopes $^1$H (circles) and $^{12}$C (squares), whereas 
open symbols represent results for $^2$H and $^{13}$C.
When comparing $E_{\rm kin}$ for $^1$H and $^{12}$C, one observes 
that the former is about twice the latter at low temperature.
This difference decreases for rising temperature.
A dashed line indicates the classical kinetic energy 
per atom, i.e., $E_{\rm kin}^{\rm cl} = 3 k_B T / 2$,
to which the quantum results converge at high $T$, irrespective
of the atomic mass. The kinetic energy for heavier isotopes 
converges faster to the classical limit.

In Fig.~4(b) we present the ratio $R$ between kinetic energies of
isotopes of the same element, carbon ($R_{\rm C}$) or hydrogen 
($R_{\rm H}$).  At low temperature, the ratio for carbon isotopes
$^{13}$C and $^{12}$C amounts to $R_{\rm C}$ = 0.959, and for 
hydrogen isotopes it is $R_{\rm H}$ = 0.712.
These values are close to those expected in a harmonic model
for the vibrational modes. In such a model, the zero-point
kinetic energy for a mode with frequency $\omega$ is given by
$E_{\rm kin}^{\rm har} = \hbar \omega / 4$.
Taking into account that the frequency
scales with the mass as $\omega \sim M^{-1/2}$, one has in
a harmonic approximation (HA) for carbon isotopes: 
$R_{\rm C} = (12 / 13)^{1/2} = 0.961$, close to the result 
derived from our PIMD simulations.
For the hydrogen isotopes, this approximation gives
$R_{\rm H} = 2^{-1/2} = 0.707$.
In the high-temperature limit, $R$ converges to unity,
since the kinetic energy approaches the classical 
value irrespective of the atomic mass. 
However, for hydrogen at $T$ = 1500 K, $R$ is still far from
unity ($R_{\rm H}$ = 0.918).

\section{Interatomic distances}

\subsection{C--C bond distance}

In this section we present results for interatomic distances 
in chair graphane.
In Fig.~5(a) we display the temperature dependence of the mean 
C--C distance, as derived from our PIMD simulations.
Results for $^{12}$C and $^{13}$C are shown as filled circles
and squares, respectively.
For the most abundant isotope, $^{12}$C, we find at low
temperature ($T \to 0$) an interatomic C--C distance of
$d_Q(0) = 1.5445$~\AA, typical of a bond between carbon atoms 
with sp$^3$ hybridization.
For $^{13}$C we obtain at low $T$ a C--C bond distance of
1.5441 \AA, somewhat smaller than that corresponding to $^{12}$C.
The statistical error bar of the bond length is 
$\sim \pm 10^{-5}$~\AA\ at $T = 1500$~K, and smaller at lower
temperatures.

\begin{figure}
\vspace{-0.6cm}
\includegraphics[width=7cm]{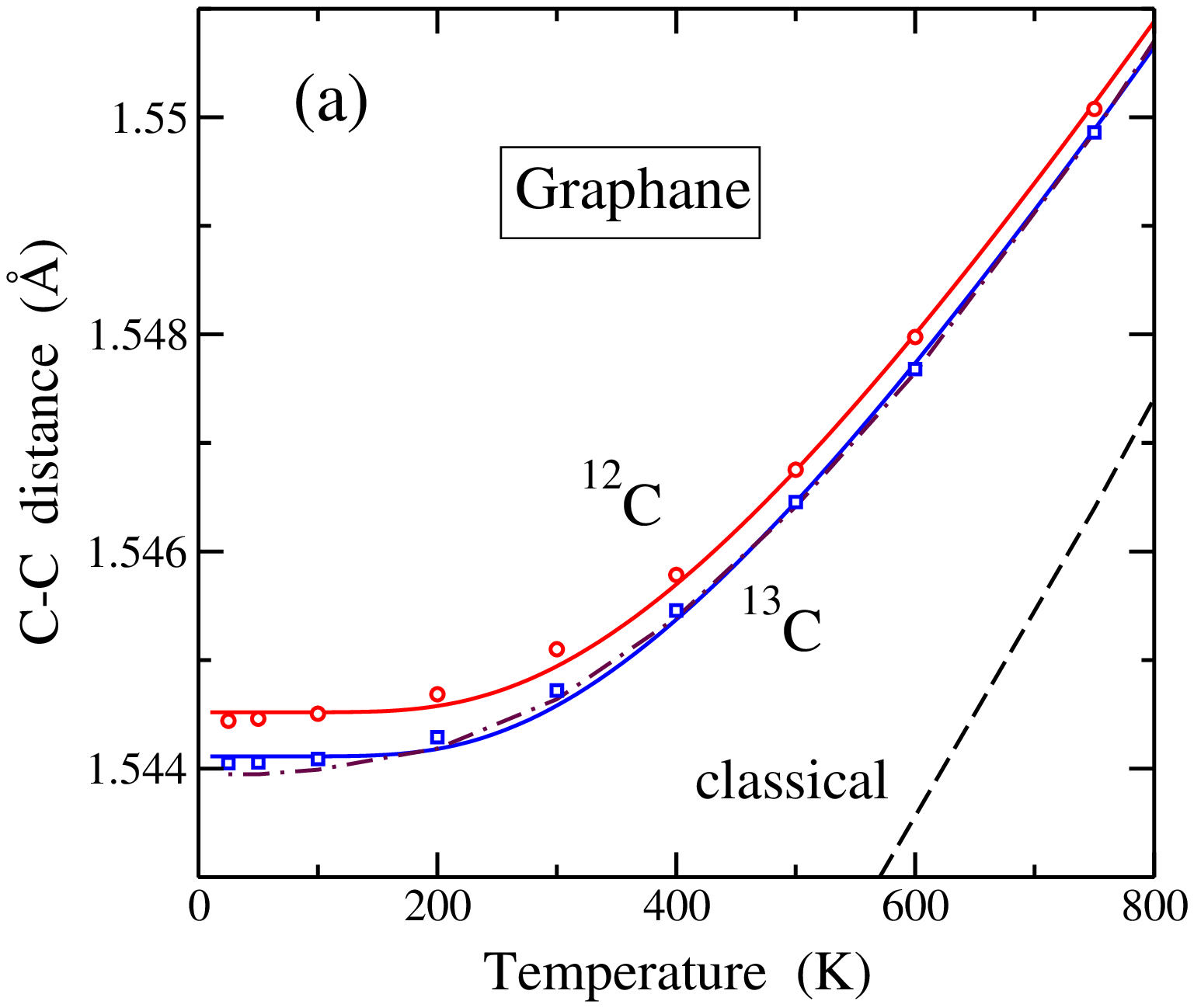}
\vspace{-0.0cm}
\includegraphics[width=7cm]{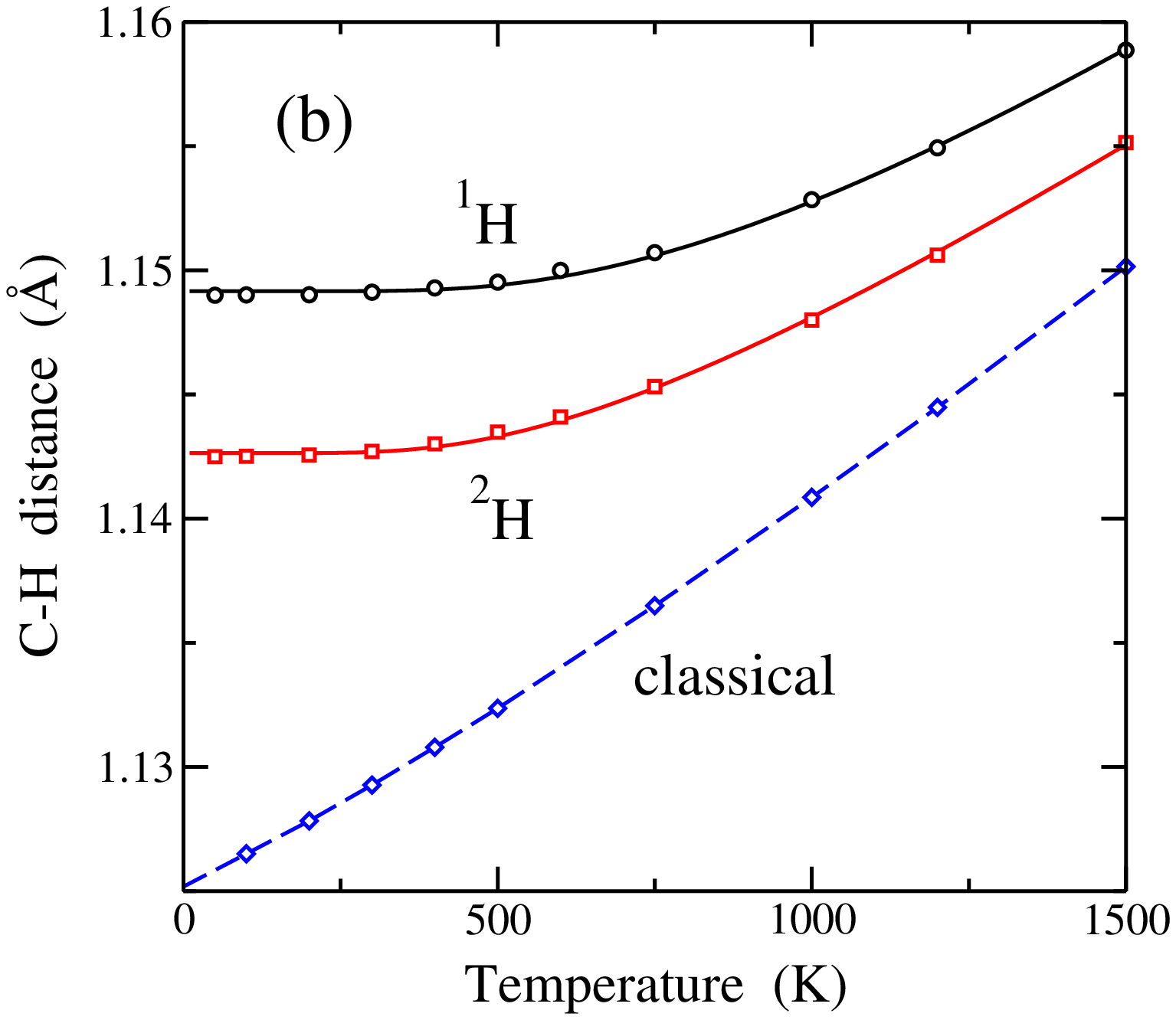}
\vspace{-0.5cm}
\caption{(a) Temperature dependence of the mean C-C distance in
$^{12}$C--$^{1}$H (circles) and $^{13}$C--$^{1}$H
graphane (squares), as derived from PIMD simulations.
A dashed line displays the classical result.
A dashed-dotted line represents the C-C bond length obtained from
PIMD simulations for $^{12}$C--$^2$H graphane.
(b) Temperature dependence of the mean C-H distance in graphane
for $^{12}$C--$^{1}$H (circles) and $^{12}$C--$^{2}$H bonds (squares).
Open diamonds are data points obtained from classical simulations.
Error bars are smaller than the symbol size.
Solid lines in (a) and (b) represent the results of the
two-parameter fitting procedure described in the text.
}
\label{f5}
\end{figure}

We also present in Fig.~5(a) the temperature dependence of the
C--C bond length derived from classical MD simulations (dashed line).
These data are independent of the isotopic mass and show an 
almost linear increase with $T$,
as can be expected for interatomic distances of crystalline 
solids in a classical approximation \cite{ki96,he01}.
These classical results for  C--C bonds in chair graphane 
converge for $T \to 0$ to $d_0$ = 1.5337 \AA, which 
corresponds to the configuration with lowest energy, $E_0$.
Note that at $T$ = 800~K the C--C bond length derived from 
classical simulations is still clearly lower than the result of
quantum simulations for $^{12}$C--$^1$H and $^{13}$C--$^1$H graphane.
This is not only due to the quantum character of carbon atoms,
but also to hydrogen atoms, whose quantum fluctuations affect
the C--C distance, as discussed below.

Zero-point vibrations of carbon atoms together with anharmonicity 
in the interatomic potential cause an expansion of the C--C bond
with respect to the classical expectancy. In the limit $T \to 0$,
this expansion for $^{12}$C amounts to 
$(\delta d)_0 = d_Q(0) - d_0 =$ 0.0108~\AA.
This means that quantum fluctuations give rise to a C--C bond dilation
of 0.7\% with respect to the classical value at low $T$. 
The difference between interatomic C--C distance for $^{12}$C
and $^{13}$C is much less than the quantum zero-point expansion
$(\delta d)_0$.  In fact, it amounts to $4 \times 10^{-4}$~\AA,
i.e., a 3.6\% of the quantum expansion.
We note that the value of $(\delta d)_0$ for C--C bonds 
in graphane is similar
to the thermal expansion predicted by the classical model from 
$T = 0$ to 650~K. Moreover, one observes in the quantum results that
the increase in C--C distance from $T$ = 0 to 300~K is rather small,
amounting to $7 \times 10^{-4}$ \AA. This value is around 
16 times smaller than the zero-point expansion $(\delta d)_0$.

To check the influence of the size of the simulation cell 
on the interatomic distances, we have verified for some selected
temperatures that our results for $N = 96$ coincide with those
obtained for $N = 216$. In fact, differences were in the order of 
the error bars of our calculated C--C distances 
(smaller than the symbols in Figs. 5(a) and 5(b)).

Our results for C--C distances derived from classical and PIMD 
simulations of chair graphane are analogous to those obtained before 
for other carbon-based materials.
For graphene, in particular, the zero-point bond expansion
$(\delta d)_0$ was found to be $8.8 \times 10^{-3}$ \AA, which 
means a 0.6\% of the bond length \cite{he16}.
For a 3D material such as diamond, the zero-point dilation of the 
C--C bond turns out to be $(\delta d)_0 = 7.4 \times 10^{-3}$~\AA,
which means a 0.5\% of the classical prediction \cite{he01}
vs a 0.7\% in graphane.

In the spirit of the QHA (see Appendix B), we present an analytic 
approximation for the temperature dependence of interatomic distances.
We assume that changes in an interatomic distance (here, C--C or C--H
bonds) are controlled by an effective vibration with frequency 
$\omega_{\rm eff}$ in the bond direction.
This effective mode is related to the bond stretching modes in
the optical phonon band LO of the crystal, but has also contributions 
of other vibrational modes for different phonon branches and 
2D wavevectors ${\bf k}$.
We express the interatomic distance at temperature $T$ as
\begin{equation}
   d_Q(T) = d_0 + \frac{\gamma_{\rm eff}}{B_{\rm eff}}  E(T)  \, ,
\label{dd0}
\end{equation}
where $d_0$ is the classical value at $T = 0$ (minimum energy
configuration),
$\gamma_{\rm eff}$ is a Gr\"uneisen parameter for the effective mode:
\begin{equation}
 \gamma_{\rm eff}  =  - \frac{d_0}{\omega_{\rm eff,0}}  
  \left. \frac{\partial \omega_{\rm eff}} {\partial d} \right|_0  \, ,
\label{geff}
\end{equation}
$B_{\rm eff}$ is a compression modulus:
\begin{equation}
 B_{\rm eff} = d_0  \left( \frac{\partial^2 E_{\rm cl} }{\partial d^2} 
         \right)_0   =  d_0  M_{\rm red}  \omega_{\rm eff,0}^2  \, ,
\label{beff}
\end{equation}
where $E_{\rm cl}$ is the potential energy of the oscillator,
given in a QHA by
$E_{\rm cl} = M_{\rm red}  \omega_{\rm eff,0}^2 (d - d_0)^2 / 2$,
$M_{\rm red}$ is the reduced mass for both atoms in the 
considered bond, and the vibrational energy at temperature 
$T$ is given by:
\begin{equation}
 E(T) =  \frac{1}{2} \hbar \omega_{\rm eff}
  \coth \left( \frac{\hbar \omega_{\rm eff}} {2 k_B T} \right)  \, .
\label{ert2}
\end{equation}
Then, we find for the interatomic distance:
\begin{equation}
  d_Q(T) = d_0 - \frac{1}{M_{\rm red} \omega_{\rm eff,0}^3} 
   \left. \frac{\partial \omega_{\rm eff}}{\partial d} \right|_0  E(T) \, ,
\label{dd0b}
\end{equation}
and the zero-point bond expansion may be written as
\begin{equation}
    (\delta d)_0 =  \frac {\hbar \gamma_{\rm eff}}
               {2 d_0  M_{\rm red} \omega_{\rm eff,0} }  \, .
\end{equation}

We have used Eq.~(\ref{dd0b}) to fit the temperature dependence of
the interatomic distances obtained from PIMD simulations of graphane.
This is a two-parameter fit, where  one variable is
the frequency $\omega_{\rm eff,0}$, and the other can be taken as the 
derivative $\partial \omega_{\rm eff} / \partial d$ or the
Gr\"uneisen parameter $\gamma_{\rm eff}$.
Values of the resulting effective frequencies and Gr\"uneisen 
parameters are given in Table~I.
The continuous lines in Fig.~5(a) represent the results of the 
fitting procedure for the C--C distance for $^{12}$C and $^{13}$C.

\begin{figure}
\vspace{-0.6cm}
\includegraphics[width=7cm]{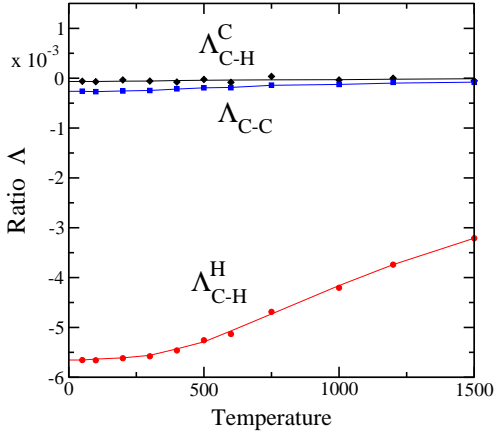}
\vspace{-0.5cm}
\caption{Temperature dependence of the ratio $\Lambda$ for
isotopic effects in interatomic distances in chair graphane.
Symbols represent results derived from PIMD simulations for
$\Lambda_{\rm C-C}$ (circles), $\Lambda_{\rm C-H}^{\rm C}$
(diamonds), and $\Lambda_{\rm C-H}^{\rm H}$ (circles).
Lines are guides to the eye.
Error bars are in the order of the symbol size.
}
\label{f6}
\end{figure}

In Fig.~6 we display the temperature dependence of the ratio
$\Lambda_{\rm C-C} = (d_Q^{13} - d_Q^{12}) / d_Q^{12}$, 
a quantitative measure of the fractional change in 
the bond length due to replacing $^{13}$C for $^{12}$C.  
For $T \to 0$, $\Lambda_{\rm C-C}$ converges to 
$-2.6 \times 10^{-4}$.
This isotopic effect is reduced as temperature rises, but it is
still observable at $T$ = 1500~K, where we find
$\Lambda_{\rm C-C} = 8(\pm 1) \times 10^{-5}$.
In graphene, this ratio was found to converge at low $T$ to
$-2.5 \times 10^{-4}$, close to our value for chair graphane. 
In the case of graphene, the so-called LCBOPII effective 
potential was used, which allowed one to consider larger simulation 
cells \cite{he20c}. The ratio $\Lambda_{\rm C-C}$ seems to be 
rather insensitive to the differences between that study of graphene 
and the present calculation for graphane (TB model, smaller cell 
sizes), so it appears as a characteristic of C--C bonds in this kind 
of 2D materials.
For comparison, we mention that in the case of
diamond, it was found at low temperature
$\Lambda_{\rm C-C} = -1.8 \times 10^{-4}$,
smaller than our data for graphane \cite{he01}.

\begin{table*}[ht]
\caption{Parameters of interatomic bonds with different isotopes.
$M_{\rm red}$: reduced mass;
 $d_0$: classical distance in the minimum-energy configuration;
$d_Q(0)$: bond length from PIMD simulations for $T \to 0$.
$\omega_{\rm eff}$ and $\gamma_{\rm eff}$ are the effective
frequency and Gr\"uneisen parameter for the different bonds.
}
%\vspace{0.5cm}
\vspace{0.2cm}
\centering
\setlength{\tabcolsep}{10pt}
\begin{tabular}{c c c c c c}
  Bond   &  $M_{\rm red}$ (amu)  &  $d_0$ (\AA)  &  $d_Q(0)$ (\AA)  &
            $\omega_{\rm eff}$ (cm$^{-1}$) &  $\gamma_{\rm eff}$
  \\ %[2mm]
\hline  \\[-4mm]
 $^{12}$C--$^{12}$C &  6.0    & 1.5337  & 1.5445 &   824  &  4.85  \\ %[2mm]

 $^{12}$C--$^{13}$C &  6.24   & 1.5337  & 1.5443 &   812  &  4.91  \\ %[2mm]

 $^{13}$C--$^{13}$C &  6.5    & 1.5337  & 1.5441 &   796  &  4.89  \\ %[2mm]

\hline  \\[-4mm]

 $^{12}$C--$^{1}$H  &  0.923  & 1.1257 &  1.1490 &  1834  &  2.64  \\ %[2mm]

 $^{12}$C--$^{2}$H  &  1.714  & 1.1257 &  1.1425 &  1372  &  2.65   \\ %[2mm]

 $^{13}$C--$^{1}$H  &  0.929  & 1.1257 &  1.1489 &  1825  &  2.64   \\ %[2mm]

 $^{13}$C--$^{2}$H  &  1.733  & 1.1257 &  1.1424 &  1362  &  2.65   \\ %[2mm]

\hline  \\[-2mm]
\end {tabular}
\label{tb:bonds}
\end{table*}

Approximate expressions for the low- and high-temperature 
isotopic effect in the interatomic distances can be obtained from
the formulation presented above in Eqs.~(\ref{dd0}) and (\ref{dd0b}).
For bonds with reduced masses $M_{\rm red}^{(1)}$ and
$M_{\rm red}^{(2)}$, we can write in the low-temperature limit:
\begin{equation}
	\Delta d_Q(0) = d_Q^{(2)}(0) - d_Q^{(1)}(0) =
        \frac {\gamma_{\rm eff}} {B_{\rm eff}}  \Delta E(0)  \, ,
\end{equation}
where $\Delta E(0) = E_2(0) - E_1(0)$ is the difference between
ground-state energies.
Taking into account Eq.~(\ref{dd0}) for the difference between classical
and quantum values for bond (1) at $T = 0$, $(\delta d)^{(1)}_0$, we find
\begin{equation}
  \Delta d_Q(0) = (\delta d)^{(1)}_0  \left[ \left( 
  1 + \frac {\Delta M_{\rm red}}{M_{\rm red}^{(1)}} \right)^{- \frac12} - 1
          \right]  \, ,
\label{ddq}
\end{equation}
with $\Delta M_{\rm red} = M_{\rm red}^{(2)} - M_{\rm red}^{(1)}$.
For $|\Delta M_{\rm red}| \ll M_{\rm red}^{(1)}$, Eq.(\ref{ddq}) can
be simplified to
\begin{equation}
	\Delta d_Q(0) = - \frac12  (\delta d)^{(1)}_0  \,
	  \frac{\Delta M_{\rm red}}{M_{\rm red}^{(1)}}  \, .
\label{ddq2}
\end{equation}
Using for $(\delta d)^{(1)}_0$ and $M_{\rm red}$ the values given 
in Table~I for C--C bonds, we find from Eq.(\ref{ddq2}) a change 
in bond distance from $^{12}$C--$^{12}$C to $^{13}$C--$^{13}$C  of
$\Delta d_Q(0) = - 4.5 \times 10^{-4}$ \AA, in 
agreement with the results of PIMD simulations for $T \to 0$.
Thus, looking at Eq.~(\ref{ddq2}), we have that this isotopic effect
(i.e., the difference $\Delta d_Q(0)$ for two isotopes)
can be directly obtained from the classical limit ($M \to \infty$) 
through the zero-point expansion $(\delta d)^{(1)}_0$.

At high temperature,
we can employ the Taylor expansion for the energy given in
Appendix A.  Then, using Eqs.~(\ref{dd0}), (\ref{beff}), and 
(\ref{ekt}) we have
\begin{equation}
 d_Q^M(T) = d_{\rm cl}(T) + \frac {\gamma_{\rm eff} \hbar^2} 
 {12 d_0 M_{\rm red} k_B T} + {\cal O} \left( (k_B T)^{-3} \right) \, ,
\label{dqdt}
\end{equation}
where we have included the contribution independent of the mass 
$M_{\rm red}$ in the classical part $d_{\rm cl}(T)$.
Using this expression, linear in $M_{\rm red}^{-1}$, we find at
$T = 750$~K a difference 
$\Delta d_Q(T) = d_Q^{13}(T) - d_Q^{12}(T) = 
-2.2 \times 10^{-4}$ \AA, which coincides with the value found
from simulations at this temperature [see Fig.~5(a)].

This discussion on the isotopic effect in the C--C bond distance
has referred to isotopic replacement of the constituent atoms of
the considered bond, i.e. $^{13}$C for $^{12}$C.
In addition to this, in the case of graphane there is also
an isotopic effect in the C--C bond distance when $^1$H is replaced 
by $^2$H, since quantum fluctuations of the C--H bond affect
the C--C bond length. Due to the light mass of hydrogen, this
effect is not negligible. In fact, from the results of our
PIMD simulations we find that substituting $^2$H for $^1$H causes
a reduction of the mean C--C bond length similar to that caused
by substituting $^{13}$C for $^{12}$C.
Results derived from PIMD for $^{12}$C--$^2$H are represented in
Fig.~5(a) as a dashed-dotted line, which can be observed at 
low $T$, and lies close to the results for $^{13}$C--$^1$H 
at high temperature.
An analytical description of this indirect isotopic effect can be
obtained from calculations based on perturbation theory, which lies
out of the scope of the present paper.

\subsection{C--H bond distance}

In Fig.~5(b) we show the mean C--H distance in graphane vs the
temperature for $^{12}$C--$^1$H (open circles) and $^{12}$C--$^2$H
isotopes (squares), as derived from our PIMD simulations. 
For comparison, we also present data for the
C--H distance derived from classical MD simulations (open diamonds).
The low-temperature quantum result for the most abundant isotope 
$^1$H is $d_Q(0)$ = 1.1490 \AA, to be compared with 
a classical value $d_0 = 1.1257$~\AA\ for $T \to 0$, which means
a zero-point bond dilation $(\delta d)_0 = 0.023$~\AA.

For the C--H bond distance, the change associated to replacing
$^2$H for $^1$H is much larger than in the data
presented in Fig.~5(a) for the C--C bond.  In fact, we find 
a reduction of the C--H distance of $6.5 \times 10^{-3}$~\AA, 
a 0.6\% of the bond length (see Table~I).
This difference between hydrogen isotopes is at low temperature  
a 27\% of the change in the C--H distance from the classical to 
the quantum model for $^1$H.
Replacing $^{13}$C for $^{12}$C causes a slight decrease in the 
C--H distance, which amount to $-7(\pm 1) \times 10^{-5}$~\AA\
at low $T$.

The temperature dependence of the interatomic distance for the
C--H bond has been fitted to the two-parameter model presented in 
Sec.~IV.A.  The continuous lines in Fig.~5(b) show the results of 
this procedure for the length of $^{12}$C--$^1$H and $^{12}$C--$^2$H 
bonds.  The respective parameters $\omega_{\rm eff}$ and
$\gamma_{\rm eff}$ are given in Table~I, along with those 
corresponding to $^{13}$C--$^1$H and $^{13}$C--$^2$H.
Note that, given the atomic species in a bond, the effective 
Gr\"uneisen parameter is independent of the isotopic masses, since
it is a magnitude calculated at the classical minimum.
The differences in $\gamma_{\rm eff}$ shown in Table~I for
different isotopes of the same species (about 1\%) may be used 
to estimate the precision of the fitting method. 
The effective frequencies $\omega_{\rm eff}$ scale with the
reduced mass of the C--H pair as $M_{\rm red}^{-1/2}$.

In Fig.~6 we have plotted the temperature dependence of the ratios 
$\Lambda_{\rm C-H}^{\rm C}$ and $\Lambda_{\rm C-H}^{\rm H}$, 
defined as:
\begin{equation}
  \Lambda_{\rm C-H}^{\rm C} = (d_Q^{13,1} - d_Q^{12,1}) / d_Q^{12,1}
\label{lambdac}
\end{equation}
and
\begin{equation}
 \Lambda_{\rm C-H}^{\rm H} = (d_Q^{12,2} - d_Q^{12,1}) / d_Q^{12,1}  \, .
\label{lambdah}
\end{equation}
These quantities, $\Lambda_{\rm C-H}^{\rm C}$ and
$\Lambda_{\rm C-H}^{\rm H}$, are the fractional changes in
the distance $d_{\rm C-H}$ under isotope changes, as indicated by the
superscripts: C for carbon and H for hydrogen isotopes.
The superscripts in the r.h.s. of Eqs.~(\ref{lambdac}) and 
(\ref{lambdah})) indicate the mass of C and H isotopes.
At low $T$, $\Lambda_{\rm C-H}^{\rm C}$ and 
$\Lambda_{\rm C-H}^{\rm H}$ converge to $-6 \times 10^{-5}$ and
$-5.7 \times 10^{-3}$, respectively.    One has 
$|\Lambda_{\rm C-H}^{\rm C}| \ll |\Lambda_{\rm C-H}^{\rm H}|$,
and in fact the former is about two orders of magnitude smaller 
than the latter. 
The isotopic ratio $\Lambda_{\rm C-H}^{\rm C}$ is close
to the sensitivity limit of our simulations.  
On the other side, $\Lambda_{\rm C-H}^{\rm H}$ is appreciably 
reduced for rising $T$ in the temperature region shown in Fig.~6, 
but at 1500~K it has a value of $-3.2 \times 10^{-3}$, 
i.e., more than one half of the low-temperature value.

To further analyze the isotopic effect in the C--H bond distance,
we use the low- and high-temperature formulas given above 
for the C--C bonds.
Thus, at low $T$, using Eq.~(\ref{ddq}) with 
$\Delta M_{\rm red} = 1.714 - 0.923 = 0.791$ and
$(\delta d)_0^{(1)}$ = 0.0233 \AA,
we find for the difference in length between
$^{12}$C-$^2$H and $^{12}$C-$^1$H bonds
$\Delta d_Q(0) = -6.2 \times 10^{-3}$ \AA, to be compared
with a difference of $-6.5 \times 10^{-3}$ \AA\ obtained from
PIMD simulations at low temperature.

At high $T$ (say 1500~ K), Eq.~(\ref{dqdt}) yields for the C--H
bonds an isotopic shift $\Delta d_Q = -3.2 \times 10^{-3}$ \AA,
somewhat smaller than that found from the simulations:
$\Delta d_Q = -3.7 \times 10^{-3}$ \AA.
This difference is not strange, taking into account the 
large effective frequency $\omega_{\rm eff}$ for C--H bonds,
which makes the series in Eqs.~(\ref{dqdt}) and (\ref{ekt}) 
to converge slower than for C--C bonds.
Moreover, anharmonicities not included in the quasiharmonic 
formulas in Eqs.~(\ref{dd0}) and (\ref{dd0b}) may show up at 
these temperatures in the presence of light species as hydrogen.

\section{Atomic motion and quantum delocalization}

\subsection{Mean-square displacements}

In this section we present the mean-square displacements
of C and H atoms in graphane, derived by PIMD simulations.
In a quantum-mechanical model, the MSD changes with the isotopic 
mass, because  the vibrational amplitudes decrease for 
increasing mass.
In a classical model, even though the vibrational frequencies 
depend on the atomic mass, the MSD for a given element
does not change with $M$.

\begin{figure}
\vspace{-0.6cm}
\includegraphics[width=7cm]{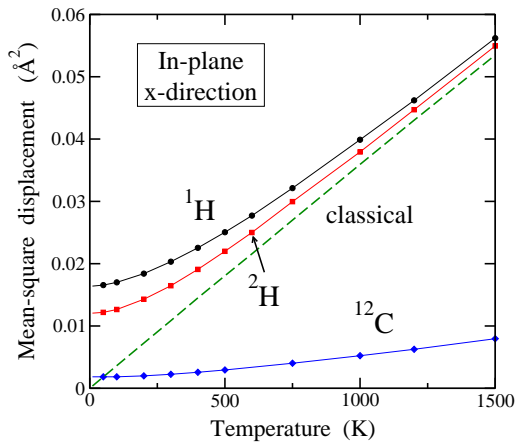}
\vspace{-0.5cm}
\caption{Mean-square displacement $(\Delta x)^2$ of
hydrogen and carbon atoms in the layer $x$-direction. Symbols
represent results of PIMD simulations for $^1$H (circles),
$^2$H (squares), and $^{12}$C (diamonds).
Results for $^{13}$C are slightly lower than those of $^{12}$C,
and are not shown for clarity of the figure.
A dashed line shows the MSD of H derived from classical MD
simulations.
}
\label{f7}
\end{figure}

In Fig.~7 we display the temperature dependence for the atomic 
MSD in the in-plane $x$-direction, $(\Delta x)^2$, 
for the hydrogen isotopes $^1$H and $^2$H, as well as for $^{12}$C.
The MSD for $^{13}$C is close to that of $^{12}$C and is not
shown for clarity of the figure.
A dashed line shows the results of classical MD simulations for
the MSD of hydrogen in the $x$-direction.
In all cases, the MSD in the in-plane $y$-direction coincides 
with the corresponding result for the $x$-direction, within the
statistical error bars of our simulations.

We remember that for a 1D harmonic oscillator of frequency $\omega$
and mass $M$, the MSD for $T \to 0$ is given by 
$(\Delta x)_0^2 = \hbar / 2 M \omega$, and $\omega \propto M^{-1/2}$, 
hence $(\Delta x)_0^2 \sim M^{-1/2}$.
This means that in a HA one expects a ratio of 0.71 for H isotopes and
0.96 for C isotopes.
From the results of our PIMD simulations we find for H a ratio
of 0.73, a little higher than that corresponding to the HA.
For C, we obtain a low $T$ a ratio of 0.96, which coincides
with the harmonic expectancy within the the precision of our 
simulations.
Given an atomic mass $M$, $(\Delta x)^2$ rises as the temperature
is increased.  Moreover, for a given $T$,
the in-plane MSD grows from the classical limit
as the atomic mass is reduced ($1/M$ rises).
Such an increase is most important at low $T$, where
quantum effects are in general more prominent. 

For atomic displacements in the out-of-plane $z$-direction,
isotopic effects are much less than for in-plane motion.
This is mainly due to the fact that, for the temperatures and
system sizes studied here, the out-of-plane motion is
dominated by the classical contribution, for which isotopic 
effects do not exist. This can be explained as follows.
The atomic MSDs derived from PIMD simulations can be divided
in two parts. One of them has classical character and is associated 
to motion of the path centroids. The other contribution is of 
quantum nature and gives a measure of the spatial extension of 
the paths \cite{he16,ha18}.
It has been found earlier that the relative quantum contribution 
to the total MSD in the $z$-direction decreases for increasing 
system size $N$ \cite{he20b,he16}  
This is due to the appearance of new low-energy vibrational modes
(small wavenumber $|{\bf k}|$) in the ZA flexural band, when 
the system size is increased.
In fact, for each size $N$ there appears a {\em crossover temperature},
$T_c$ above which the classical part is the main contribution to the
MSD in the $z$-direction, and $T_c$ goes down for rising $N$.
For the system size considered here ($N = 96$), this crossover
temperature is around 100~K for H and much less for C, 
so the MSD in the $z$-direction is dominated by the classical 
contribution in almost the whole temperature range of our simulations. 
In summary, quantum effects and thereby isotopic effects in
the out-of-plane direction are much less important than in the 
$(x,y)$ plane, and cannot be precisely obtained given the 
accuracy of our calculations.

\subsection{Tilt of the C--H bonds}

We study the orientation of the C--H bonds by using
spherical coordinates $(\theta,\phi)$.
$\theta$ is the polar angle between the $z$-direction and the 
C--H bond, and $\phi$ is the azimuth measured on the $(x,y)$ plane.
In the lowest-energy configuration one has $\theta = 0$, which
means that the C--H bonds are exactly perpendicular to the 
$(x,y)$ plane.
The temperature dependence of the mean polar angle, 
$\langle \theta \rangle$, is shown in Fig.~8 for the hydrogen
isotopes $^1$H (squares) and $^2$H (diamonds), as derived from
PIMD simulations. The results of classical simulations are
displayed as solid circles.
In the classical approach, $\langle \theta \rangle$ vanishes
for $T \to 0$, and increases for rising temperature as
$\sqrt{T}$.

\begin{figure}
\vspace{-0.6cm}
\includegraphics[width=7cm]{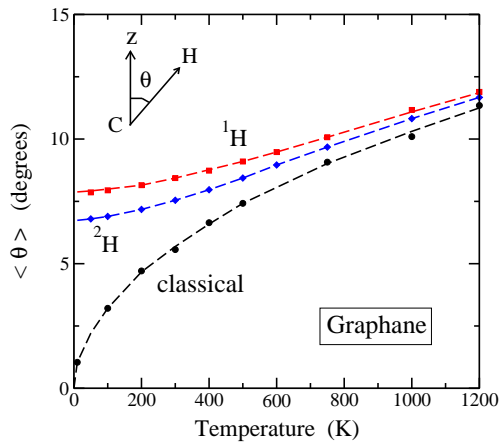}
\vspace{-0.5cm}
\caption{Mean polar angle $\langle \theta \rangle$ between the
C-H bond and the $z$-direction vs temperature.
Solid squares and diamonds represent results of PIMD simulations
for $^{12}$C--$^1$H and $^{12}$C--$^2$H graphane, respectively.
Circles indicate the mean angle $\langle \theta \rangle$
obtained from classical simulations.
Lines are guides to the eye.
Error bars are in the order of the symbol size.
}
\label{f8}
\end{figure}

The results of PIMD simulations reach a finite value in the
low-temperature limit.  For $^1$H and $^2$H we find
$\langle \theta \rangle_0$ = 7.8 deg and 6.7 deg, respectively.
Although $\langle \theta \rangle_0$ is somewhat lower for
$^2$H, the angle dispersion caused by atomic zero-point motion
is appreciable for both H isotopes.
In each case, $\langle \theta \rangle$ increases as temperature is
raised, and at $T$ = 1500 K it takes a value close to 13.0 deg
for both isotopes. At this temperature the classical value is
slightly lower than the results of PIMD simulations.

The dispersion in the polar angle $\theta$ can be analyzed by
defining the probability distribution $P(\theta)$, which
verifies the normalization condition
\begin{equation}
  \int_0^{\pi}  P(\theta) \sin \theta \, d \theta  = 1    \, .
\end{equation}
Our results of both quantum and classical simulations
show that $P(\theta)$ follows a Gaussian distribution within
the statistical noise:
\begin{equation}
    P(\theta) = c \, \exp(- a \theta^2) \, ,
\label{ptheta}
\end{equation}
$c$ being a normalization constant defined as
\begin{equation}
   c^{-1} = \int_0^{\pi} \sin \theta \, 
             \exp(- a \theta^2) \, d \theta  \, .
\end{equation}
The width of the Gaussian distribution is given by the parameter 
$a$ in Eq.~(\ref{ptheta}). The distribution $P(\theta)$ becomes 
wider as temperature is raised and the angular dispersion increases,
with a corresponding decrease in the parameter $a$.
From our PIMD simulations, we obtain for $^1$H:
$a$ = 40.6 and 14.8 rad$^{-2}$ for $T$ = 100 and 1500 K,
respectively.
In the classical limit, the Gaussian distribution $P(\theta)$
converges to a Dirac $\delta$-function for $T \to 0$, and
$a$ diverges to infinity.

Given the Gaussian distribution $P(\theta)$ in Eq.~(\ref{ptheta}),
the mean polar angle $\langle \theta \rangle$ may be written as
\begin{equation}
  \langle \theta \rangle =   \int_0^{\pi} 
        P(\theta) \, \theta \,  \sin \theta \, d \theta  \; .
\end{equation}
In the low-temperature limit we find $a_0$ = 42.2 and 
57.3 rad$^{-2}$ for $^1$H and $^2$H, respectively.
These values yield for the mean polar angle:
$\langle \theta \rangle_0$ = 0.136 rad = 7.8 deg for $^1$H and 
$\langle \theta \rangle_0$ = 0.117 rad = 6.7 deg for $^2$H,
in agreement with the mean values derived directly from the
simulations.
Note that the ratio between the squared mean angles
$\langle \theta \rangle_0^2$ for $^2$H and $^1$H is 0.74,
somewhat higher than the ratio $1 / \sqrt{2} = 0.71$ expected for
small angle fluctuations in a HA. 
This indicates that the accuracy of a harmonic calculation is
reduced for the relatively large amplitude of the polar
angle oscillations, caused by the light mass of hydrogen.

\section{Area of the graphane sheet}

The temperature dependence of the in-plane area $A_p$ is displayed
in Fig.~9. Symbols represent results of PIMD simulations for
$^{12}$C--$^1$H (circles) and $^{13}$C--$^1$H graphane (squares).
A dashed line on the right side of the figure shows the in-plane 
area found in classical simulations, which converges to the
quantum results at high temperature.
For $^{12}$C--$^1$H graphane, we find at low temperature
$A_p$ = 2.8041~\AA$^2$/(C atom), to be compared with 
the classical limit $A_0 = 2.7758$~\AA$^2$/(C atom) at 
$T = 0$ (not shown in the figure).
This means a zero-point increase in the area of 
$(\delta A_p)_0 = 0.028$~\AA$^2$/(C atom).
The difference between classical and quantum results decreases
for rising $T$, as quantum effects turn less relevant.
At the highest temperature shown in Fig.~9, $T = 1000$~K,
it amounts to  $9 \times 10^{-3}$ \AA$^2$/(C atom)
Note that we consider for $A_p$ the isotopic effect associated
to the C mass, since C--C bonds define the graphane network
in the layer plane, and essentially control the changes in
the in-plane area.

\begin{figure}
\vspace{-0.6cm}
\includegraphics[width=7cm]{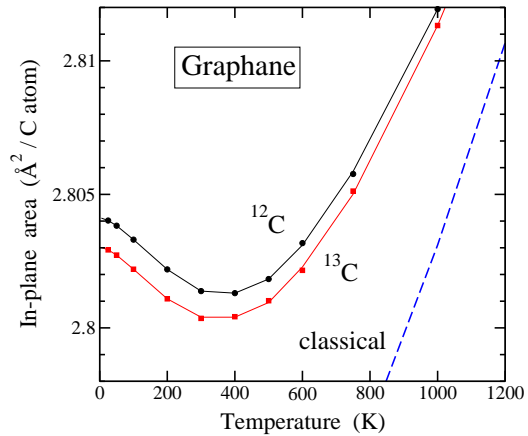}
\vspace{-0.5cm}
\caption{Temperature dependence of the in-plane area $A_p$
of $^{12}$C--$^1$H and $^{13}$C--$^1$H graphane, as derived
from PIMD simulations for $N = 96$.
Solid lines are guides to the eye.
Error bars are in the order of the symbol size.
A dashed line indicates the result of classical MD simulations.
}
\label{f9}
\end{figure}

In the PIMD results we find a decrease in in-plane area
in the temperature range from $T = 0$ to $T \sim 400$~K, which
is not found in classical simulations \cite{he20b}.
This in-plane area contraction, i.e., $d A_p / d T < 0$ at low $T$, 
found in quantum simulations, is caused by out-of-plane atomic motion,
which overshadows the thermal expansion of C--C bonds.
At $T > 400$~K, the bond dilation dominates over the shrinkage of
the in-plane area due to atomic vibrations in the $z$-direction, and
$d A_p / d T > 0$.  The decrease in $A_p$ below $T = 400$~K does not 
appear in classical simulations, since the relative contributions 
of the in-plane and out-of-plane vibrational modes are not adequately
described at low $T$.
Moreover, an important outcome of the quantum simulations is that 
the resulting area $A_p$ verifies $d A_p / d T \to 0$ in the low-$T$ 
limit, in line with the third law of Thermodynamics \cite{ca60}.
We find this to happen for graphane, as was also found before
for graphene \cite{he16}. Classical simulations, however,
fail to fulfill this thermodynamic requirement at low temperature.

Huang {\em at al.} \cite{hu13b,hu14} calculated several thermodynamic 
and vibrational properties of graphane by using density-functional 
perturbation theory. The results presented by these authors for
the in-plane area (or lattice constant) up to $T = 800$~K are 
similar to those given here. In particular, they found a minimum 
for the lattice constant (vanishing expansion coefficient) at 
$T \approx 400$~K, as happens in our results for the in-plane area.

\begin{figure}
\vspace{-0.6cm}
\includegraphics[width=7cm]{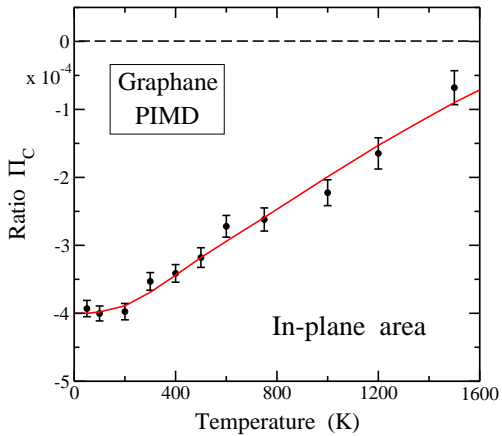}
\vspace{-0.5cm}
\caption{Temperature dependence of the ratio
$\Pi_{\rm C} = (A_p^{13} - A_p^{12}) / A_p^{12}$,
derived from PIMD simulations of chair graphane.
(circles). The solid line is a guide to the eye.
}
\label{f10}
\end{figure}

For a given temperature, our results indicate that the area $A_p$ 
decreases as the carbon isotopic mass is raised.
At $T = 50$~K, the shift in $A_p$ associated to replacing $^{13}$C 
for $^{12}$C is $\Delta A_p = -1.1 \times 10^{-3}$ \AA$^2$/(C atom),
In Fig.~10 we show the temperature dependence of the ratio
$\Pi_{\rm C} = (A_p^{13} - A_p^{12}) / A_p^{12}$, where $A_p^M$ is
the in-plane area for graphane with carbon isotopic mass $M$.
This ratio is found to converge at low $T$ to
$\Pi_{\rm C} = -3.9(1) \times 10^{-4}$. 
At $T = 1500$~K, this isotopic effect
is still observable, as $\Pi_{\rm C} < 0$, even taking into account 
the relatively large error bar at this temperature.   Note that 
the error bars for $\Pi_{\rm C}$ are larger than for the ratio 
$\Lambda$ presented in Fig.~6 for C--C and C--H bonds, which
are not shown as they are in the order of the symbol size.
The larger statistical noise in $A_p$, as compared to interatomic
distances, is caused in part by its larger fluctuations at a given 
temperature (see Fig.~2).
Such fluctuations of $A_p$ are due to slow out-of-plane
bending modes with large wavelengths (small $|{\bf k}|$)
and low vibrational frequencies.
Moreover, the statistics for interatomic distances is better
than that of $A_p$ because in a simulation step one has a
value for the in-plane area vs $3 N / 2$ and $N$ values for 
C--C and C--H distances, respectively.

Our results for the isotopic effect on the in-plane area can be 
understood from the trends calculated in a QHA for the vibrational 
modes \cite{de96,ga96,mo05}.
In this approximation, the frequencies $\omega_r({\bf k})$ 
($r$: index for the phonon bands in graphane) are supposed
to change with the area $A_p$, and for a given $A_p$ the modes
are considered to be harmonic.
The difference $\Delta A_p(0) = A_p^{M_2}(0) - A_p^{M_1}(0)$
for isotopic mass $M_2$ and $M_1$ can be written for $T \to 0$ as
(see Appendix B):
\begin{equation}
 \Delta A_p(0) = - \frac{C}{2 M_1^{3/2}}  \Delta M    \,  ,
\label{dap0b}
\end{equation}
where $\Delta M  = M_2 - M_1$, and $C$ is a constant independent 
of the mass.

In the low-temperature limit, the increase in $A_p$ due to
quantum fluctuations is given from Eq.~(\ref{ap2}):
\begin{equation}
      A_p^{M}(0) - A_0 = \frac{C}{M^{1/2}}   \, .
\label{ap1b}
\end{equation}
Using Eq.~(\ref{ap1b}), we may eliminate the constant $C$ in
Eq.~(\ref{dap0b}), and the difference $\Delta A_p(0)$ for two isotopes 
can be expressed as
\begin{equation}
  \Delta A_p(0) = - \frac12  \left[ A_p^{M_1}(0) - A_0  \right]  
       \frac{\Delta M}{M_1}   \, .
\label{dap0c}
\end{equation}
Thus, the isotopic effect in the area $A_p$ at low temperature
can be obtained from the zero-point expansion 
$(\delta A_p)_0 = A_p^{M}(0) - A_0$, 
i.e. the increase in area from the classical limit to the quantum 
result for a given isotopic mass, e.g. $M_1$.
For carbon mass $M_1 = 12$~amu, we have a zero-point increase of 
0.028~\AA$^2$/(C atom), so we find from Eq.(\ref{dap0c}) a
difference $\Delta A_p(0) = -1.17 \times 10^{-3}$ \AA$^2$/(C atom)
between $^{13}$C and $^{12}$C graphane.
This value is very close to the difference obtained from the 
results derived directly from the PIMD simulations for both isotopes,
shown in Fig.~9. 
Note the similarity between Eq.~(\ref{dap0c}) for $\Delta A_p(0)$ 
and Eq.~(\ref{ddq2}) for $\Delta d_Q(0)$, 
where the role of the isotopic mass in the former
is played by the reduced mass $M_{\rm red}$ in the latter.

The relative change in $A_p$ found here for isotopic substitution in
graphane at low $T$ is close to that obtained earlier 
for graphene \cite{he20c}.
As indicated above, in the case of graphene an effective potential
(the so-called LCBOPII) was employed to describe the interatomic
interactions, different in nature from the TB model used here.
Moreover, in the case of graphene it was feasible to deal with 
simulation cells including thousands of carbon atoms, which
is prohibited for the computationally more exigent TB model.
Taking into account these differences, it seems that the magnitude 
of the isotopic effect in the in-plane area is a common characteristic
of 2D carbon-based materials.

In this section we have considered the area $A_p$ for
isotopically pure graphane, i.e., made up of $^{12}$C--$^1$H or 
$^{13}$C--$^1$H. 
For mixtures of carbon isotopes, one expects an in-plane area
corresponding to a linear interpolation between those found
for isotopically pure samples, as happens for the volume of
3D materials.
We have checked this question by carrying out PIMD simulations
for carbon mean mass of 12.25, 12.5, and 12.75 amu.
In each case, two types of simulations were performed. 
In the first one, we considered simulation cells with adequate 
proportions of $^{12}$C and $^{13}$C to give 
the corresponding mean mass.
In the second case, we took cells where each carbon nucleus 
in graphane has a mass equal to the average mass.
This is the so-called {\em virtual-crystal approximation}
\cite{de96,he99,ca05b,he09c}.
We found that for a given carbon mean mass, both kinds of
simulations yield the same in-plane area, within the numerical 
precision of our calculations.
This means the validity of the virtual-crystal approximation
for calculating areas of 2D materials, as was found before
for volumes or lattice parameters of 3D crystals.
Moreover, the results for $A_p$ in isotopically mixed graphane
agree with a linear interpolation of data obtained for
isotopically pure samples.

To end this section, we emphasize that the isotopic effects 
studied here are certainly larger than the sensitivity of 
several diffraction methods.
For example, x-ray standing waves \cite{ka98} and Bragg 
backscattering measurements \cite{wi02} were shown
to accurately yield the magnitude of isotopic effects in 
lattice parameters and interatomic distances of crystalline
solids. For lattice parameters, in particular, the uncertainty 
of these methods in $\Delta a / a$ can be less than $10^{-6}$.

\begin{table}[ht]
\caption{C--C bond distance and area $A_p$ of
$^{12}$C--$^1$H graphane, as derived from PIMD simulations
at $T =$ 300 and 500 K for various in-plane stresses $\tau$.
}
\vspace{0.2cm}
\centering
\setlength{\tabcolsep}{10pt}
\begin{tabular}{c c c c}
 $T$ (K) & $\tau$ (eV/\AA$^2$) & $d_Q^{12}$ (\AA) & $A_p$ (\AA$^2$/C atom)
  \\ %[2mm]

\hline  \\[-4mm]
   300  &   -0.4  &   1.57945    &  2.95441   \\  %[2mm]
   300  &   -0.2  &   1.56071    &  2.87178   \\  %[2mm]
   300  &    0.0  &   1.54509    &  2.80138   \\  %[2mm]
   300  &    0.2  &   1.53180    &  2.73457    \\  %[2mm]

\hline  \\[-4mm]
   500  &   -0.4  &   1.58183    &  2.95890   \\  %[2mm]
   500  &   -0.2  &   1.56266    &  2.87411   \\  %[2mm]
   500  &    0.0  &   1.54676    &  2.80186   \\  %[2mm]
   500  &    0.2  &   1.53332    &  2.73236    \\  %[2mm]

 \hline  \\[-2mm]
\end {tabular}
\label{tb:stress}
\end{table}

\section{Pressure effects}

In this section we analyze the stress dependence of the isotopic
effect in the in-plane area $A_p$ and C--C bond distance. 
The in-plane stress $\tau$ is expected to affect mainly these bonds,
since such an stress is perpendicular to C--H bonds in graphane. 
In Table~II we present results for the C--C bond distance and
in-plane area $A_p$ for $^{12}$C--$^1$H graphane under various 
stresses $\tau$ at $T$ = 300 and 500 K. 
Both $d_Q^{12}$ and $A_p$ increase or decrease for
rising tensile ($\tau < 0$) or compressive ($\tau > 0$) stress, 
respectively.

From the stress derivative of $A_p$ at $\tau = 0$, one can 
calculate the 2D modulus of hydrostatic compression,
$B_p = - A_p / (\partial A_p / \partial \tau)$ (inverse of the 
in-plane compressibility) \cite{be96b}.  From the results of our 
simulations we find for $^{12}$C--$^1$H graphane at 300 K:
$B_p = 8.4(2)$ eV/\AA$^2$. Although $B_p$ for $^{13}$C--$^1$H 
graphane is expected to be larger \cite{cl17}, our result in this 
case is indistinguishable from that for $^{12}$C--$^1$H, 
taking into account the associated error bar.
This is mainly due to the relatively large fluctuations in 
$A_p$, as discussed in Sec.~VI.
At $T$ = 500 K, our simulations yield $B_p = 8.2(2)$ eV/\AA$^2$
for $^{12}$C and $^{13}$C graphane.

\begin{figure}
\vspace{-0.6cm}
\includegraphics[width=7cm]{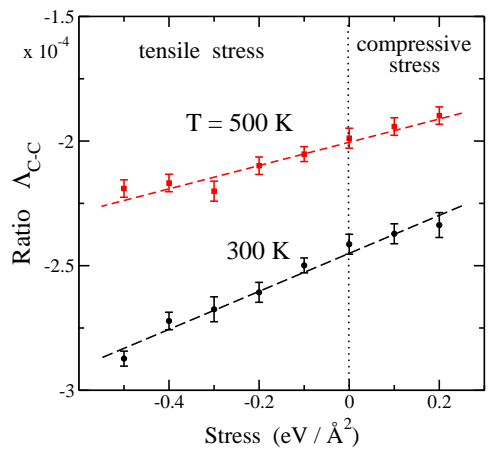}
\vspace{-0.5cm}
\caption{Stress dependence of
$\Lambda_{\rm C-C} = (d_Q^{13} - d_Q^{12}) / d_Q^{12}$,
as obtained from PIMD simulations of chair graphane at
$T = 300$~K (circles) and 500 K (squares).
Dashed lines are linear fits to the data points.
A vertical dotted line separates the regions of tensile
($\tau < 0$) and compressive stress ($\tau > 0$).
}
\label{f11}
\end{figure}

Differences between both carbon isotopes are more clearly 
observed in the stress dependence of the C--C bond distance.
This is presented in Fig.~11, where we display the relative
change of the C--C distance, $\Lambda_{\rm C-C}$,
at $T$ = 300 K (circles) and 500 K (squares).
For a given temperature, we observe that the isotopic shift becomes
larger ($|\Lambda_{\rm C-C}|$ increases) as the tensile stress grows, 
and is reduced for rising compressive stress.
At 300 K, the parameter $\Lambda_{\rm C-C}$ changes by a 20\% from 
unstressed graphane to a tensile stress of $\tau = -0.5$~eV/\AA$^2$.
At 500 K, the corresponding change is about 10\%.
The results of our PIMD simulations show a linear dependence 
of $\Lambda_{\rm C-C}$ vs $\tau$ in the region of
in-plane stresses presented in Fig.~11.
We find that the slope $\partial \Lambda_{\rm C-C} / \partial \tau$ 
decreases as temperature is raised.
In fact, at $T$ = 300 and 500 K, we obtain slopes of 
7.6 and $4.6 \times 10^{-5}$~\AA$^2$/eV, respectively.

We now look for an explanation of the positive sign of the
derivative $\partial \Lambda_{\rm C-C} / \partial \tau$ found
from our PIMD simulations of chair graphane. Applying a tensile 
stress gives rise to a softening of the in-plane acoustic and 
optical phonons (positive Gr\"ueisen parameters), which should 
mainly contribute to the effective frequency $\omega_{\rm eff}$.
Then, a decrease in $\omega_{\rm eff}$ causes a reduction
in its associated zero-point vibrational energy, and consequently
in the difference $d_Q^{12} - d_Q^{13}$ [see Eq.~(\ref{dd0})]. 
One could argue that this
should correspond to a reduction in $|\Lambda_{\rm C-C}|$,
contrary to the results of our simulations displayed in Fig.~11.
The key point is that a decrease in $d_Q^{12} - d_Q^{13}$ does not
necessarily lower $|\Lambda_{\rm C-C}|$, as shown below.

The positive sign of $\partial \Lambda_{\rm C-C} /\partial \tau$
can be understood from the low-$T$ stress derivatives of the
interatomic distances $d_Q^{12}$ and $d_Q^{13}$.
From Eq.~(\ref{dd0}) we have at low temperature:
\begin{equation}
  d_Q^M(0) = d_0 + \frac{Z}{d_0 \omega_{\rm eff} M_{\rm red} }  \, ,
\label{dm0}
\end{equation}
with the parameter $Z = \hbar \gamma_{\rm eff} / 2$,
independent of the stress and the isotopic mass.
In Eq.~(\ref{dm0}), both the classical low-$T$ distance $d_0$ and 
the effective frequency $\omega_{\rm eff}$ change with
the applied in-plane stress $\tau$.
Thus, we have for isotopic mass $M$:
\begin{equation}
 \frac {\partial d_Q^M(0)}{\partial \tau} = 
   \frac {\partial d_0}{\partial \tau} - 
   \frac{Z}{d_0 \omega_{\rm eff} M_{\rm red}}
   \left( \frac{1}{d_0} \frac {\partial d_0}{\partial \tau} + 
   \frac{1}{\omega_{\rm eff}} \frac {\partial \omega_{\rm eff}}
          {\partial \tau}  \right)   \, .
\label{partd}
\end{equation}
This expression can be written as
\begin{equation}
  \frac {\partial d_Q^M(0)}{\partial \tau} = 
    \frac {\partial d_0}{\partial \tau} - 
    \frac{Z'}{d_0 \omega_{\rm eff} M_{\rm red}}
\label{partd2}
\end{equation}
with the parameter $Z' > 0$ and independent of $M_{\rm red}$
(see Appendix~C).    Thus, we have
\begin{equation}
 \frac {\partial d_Q^{12}(0)}{\partial \tau} -
 \frac {\partial d_Q^{13}(0)}{\partial \tau} < 0  \, ,
\label{partd3}
\end{equation}
and the difference $d_Q^{12} - d_Q^{13}$ decreases under a
compressive stress ($\tau > 0$).

The stress derivative of the parameter $\Lambda_{\rm C-C}$ 
may be written for low temperature as
\begin{equation}
  \frac{\partial \Lambda_{\rm C-C}}{\partial \tau} = 
          \frac{F_0}{d_Q^{12}(0)^2}  \, ,
\label{dlambda}
\end{equation}
where $F_0$ is a function of the distances $d_Q^M(0)$ and
their derivatives $\partial d_Q^M(0) / \partial \tau$
for $M$ = 12 and 13 amu.
Using the expressions for the interatomic distances and their
derivatives given in Eqs.~(\ref{dm0}) and (\ref{partd2}),
one finds
\begin{equation}
  F_0 =  Z''(\gamma_{\rm eff} - 2)   \, ,
\end{equation}
with the parameter $Z'' > 0$ (see Appendix C).
$F_0$ is positive for $\gamma_{\rm eff} > 2$, as occurs
in our case for C--C bonds in graphane, and consequently
$\partial \Lambda_{\rm C-C} / \partial \tau > 0$. 

A positive slope for $\Lambda_{\rm C-C}$ vs $\tau$ is also
found from the PIMD simulations of graphane at finite
temperatures, as shown in Fig.~11 for $T =$ 300 and 500~K.
The slope $\partial \Lambda_{\rm C-C} /\partial \tau$
decreases for rising $T$ and eventually vanishes in the
high-$T$ limit, where the difference $d_Q^{12} - d_Q^{13}$ and
its stress derivative vanish (classical limit).

\section{Summary}

The PIMD technique is a versatile method to study isotopic 
effects in molecular systems and solids.
For crystalline solids, in particular, this procedure
enables one to consider structural and phonon-related properties,
further than standard approaches based on a harmonic
approximation for the lattice vibrational modes.
The atomic mass is an input parameter in the simulations,
so one can examine the influence of the isotopic mass
of the constituent atoms on the physical properties of 
the material, in our case a 2D crystalline solid as 
chair graphane. 

From an analysis of the internal and kinetic energy, as well
as from the atomic mean-square displacements, we find that
our low-temperature results are compatible with a harmonic
approximation for the vibrational modes, i.e., relations of 
these variables for different isotopic masses are close
to those expected in such an approximation.
However, structural variables change for
different isotopes, since they are sensitive to the anharmonicity
of the interatomic potential through the atomic zero-point 
motion. Such anharmonicity is quantitatively detected by 
PIMD simulations at relatively low temperatures.

We have analyzed the variation of interatomic bond distances 
and in-plane area $A_p$ with the isotopic mass of carbon and
hydrogen in a temperature range from $T$ = 50 to 1500 K.
At low temperature, the fractional change in the length of
$^{13}$C--$^{13}$C bonds with respect to $^{12}$C--$^{12}$C 
bonds in graphane is found to be 
$\Lambda_{\rm C-C} = -2.6 \times 10^{-4}$. 
For the in-plane area $A_p$, the corresponding relative 
change is $\Pi_{\rm C} = -3.9 \times 10^{-4}$.
These isotopic effects are reduced as temperature is increased,
and at 1000~K we obtain relative variations of
$-1.2$ and $-2.2 \times 10^{-4}$
for the C--C distance and the layer area, respectively.
Much larger is the fractional change in C--H bond distance 
caused by replacing $^2$H for $^1$H. At low temperature
we find $\Lambda_{\rm C-H}^{\rm H} = -5.7 \times 10^{-3}$.
At $T$ = 1000~K, this parameter is still relatively large,
i.e., $-4.2 \times 10^{-3}$.

On the basis of a QHA for the vibrational modes,
we have shown that the isotopic effect in the in-plane area $A_p$ 
at low $T$ may be derived from the zero-point expansion
$(\delta A_p)_0 = A_p^{M}(0) - A_0$,
i.e. from the rise in area from the classical limit to the quantum
value for a reference isotopic mass $M$. This fact derived from
the QHA agrees with the results of our PIMD simulations.

In the same spirit of the QHA, the temperature dependence of 
interatomic distances can be understood as caused by
an effective vibration with frequency
$\omega_{\rm eff}$ in the bond direction. 
The dependence on $T$ of C--C and C--H bond distances, expressed
by Eqs.~(\ref{dd0}) and (\ref{dd0b}), is in line with the data
yielded by PIMD simulations [see Fig.~5].
This approach provides us with a consistent picture to understand
isotopic effects in the bond lengths in graphane.

We have also considered the influence of an external in-plane stress 
$\tau$ on the isotopic effect (parameter $\Lambda_{\rm C-C}$ ) 
in graphane.
We find that the magnitude of $\Lambda_{\rm C-C}$ is
increased for a tensile stress ($\tau < 0$) and
reduced for a compressive one ($\tau > 0$).
This is in line with a decrease in $\omega_{\rm eff}$ for
$\tau < 0$ and a rise for $\tau > 0$.  \\  \\

\begin{acknowledgments}
This work was supported by Ministerio de Ciencia e Innovaci\'on
(Spain) through Grant PGC2018-096955-B-C44.
\end{acknowledgments}

%  -----------------------------------------------------------------

\appendix

\section{Inverse-mass expansion for a harmonic oscillator}

For a 1D harmonic oscillator with spring constant $k$,
the potential energy is given by $V(x) = k x^2 / 2$.
The energy at temperature $T$ is:
\begin{equation}
   E = \frac12 \hbar \omega \coth \left(  
          \frac{\hbar \omega}{2 k_B T}  \right)  \, ,
\end{equation}
with the frequency $\omega = \sqrt{k/M}$, and $M$ the mass. 
The energy can be written as a function of the inverse mass
$x = 1 / M$ as
\begin{equation}
  E =  A \sqrt{x} \coth (B \sqrt{x})  \, ,
\end{equation}
with $A = \hbar \sqrt{k} / 2$ and $B = A / k_B T$.

A Taylor expansion of the energy for small $x$ yields:
\begin{equation}
  \frac{E}{A} =  \frac1B + \frac13 B x - \frac{1}{45} B^3 x^2 + ...
\label{ea}
\end{equation}
or
\begin{equation}
   E = k_B T + \frac{\hbar^2 k}{12 \, k_B T} \, x - 
       \frac{\hbar^4 k^2}{720 \, (k_B T)^3} \, x^2 + ...
\label{ekt}
\end{equation}
The first term in this expansion, independent of the mass, is
the classical thermal energy at temperature $T$.
The expansion in Eqs.~(\ref{ea}) and (\ref{ekt}) converges 
for $B \sqrt{x} < \pi$, or $T > \hbar \omega / 2 \pi k_B$.

The linear term in $x$, with the prefactor
$A B / 3 = \hbar^2 k / 12 k_B T$ diverges for $T \to 0$ and
the expansion breaks down. 
In fact, at $T = 0$ the energy is given by
\begin{equation}
  E(0) = \frac12 \hbar \omega = \frac12 \hbar 
           \left( \frac{k}{M} \right)^{\frac12}  \, ,
\end{equation}
and the energy $E(0) \sim M^{-1/2}$ is not an analytic function 
of the inverse mass,
so that a Taylor expansion is not possible.

Defining $\lambda = B^2 x$, we have
\begin{equation}
 \frac{E}{A} =  \frac1B + \frac{1}{3B} \lambda - 
                \frac{1}{45 B} \lambda^2 + ...
\end{equation}
or
\begin{equation}
 E =  k_B T  \left( 1 + \frac13 \lambda - \frac{1}{45} \lambda^2 + ...
             \right)   \, ,
\label{e_lambda}
\end{equation}
with
\begin{equation}
    \lambda = \left( \frac {\hbar \omega}{2 k_B T} \right)^2  \, .
\end{equation}

For small $\lambda$, we use the first-order approximation
\begin{equation}
   E  \approx  k_B T +  \frac13 k_B T \lambda  \, ,
\end{equation}
valid for $\lambda \ll 1$, or
\begin{equation}
  k_B T  \gg  \frac12 \hbar \left( \frac{k}{M} \right)^{\frac12} = 
	\frac12 \hbar \omega   \, .
\end{equation}

Note that a linear approximation $E =  k_B T (1 + \lambda / 3)$
is very accurate for many purposes, even for 
$\hbar \omega \sim k_B T$. In fact, for
$\hbar \omega = k_B T$ (i.e., $\lambda = 1/4$) the second-order 
term in the r.h.s. of Eq.~(\ref{e_lambda}) is 60 times
smaller than the first-order one.

\section{Quasiharmonic approximation}

In a QHA for 2D crystalline solids, the vibrational modes are considered
as harmonic oscillators with frequencies $\omega_r({\bf k})$, which 
depend on the in-plane area of the material \cite{de96,mo05,he20c}. 
The index $r$ indicates the phonon bands (12 in the case of graphane),
and ${\bf k}$ is the wavevector in the 2D hexagonal Brillouin zone of 
the reciprocal lattice \cite{ra19}.

The equilibrium area $A_p$ for isotopic mass $M$ at 
temperature $T$ and external stress $\tau = 0$ can be obtained by 
minimizing the Helmholtz free energy of the system \cite{mo05,he20c}.
This gives
\begin{equation}
A_p^M(T) = A_0 + \frac{1}{N B_0}
   \sum_{r, {\bf k}}  \gamma_r({\bf k})  E_r({\bf k},T)
    \hspace{0.2cm}  ,
\label{apt}
\end{equation}
where $E_r({\bf k},T)$ is the vibrational energy of mode $r, {\bf k}$:
\begin{equation}
 E_r({\bf k},T) =  \frac{1}{2} \hbar \omega_r({\bf k}) 
  \coth \left( \frac{\hbar \omega_r({\bf k}) } {2 k_B T} \right)  \, .
\label{ert}
\end{equation} 
In Eq.~(\ref{apt}),
$A_0$ is the in-plane area for $T = 0$ and $M \to \infty$
(classical limit),   $B_0$ is the 2D modulus of hydrostatic 
compression for the minimum-energy configuration \cite{be96b}, and
\begin{equation}
 \gamma_r({\bf k}) = - \left. \frac {\partial \ln \omega_r({\bf k})}
   {\partial \ln A_p} \right|_0
\end{equation}
is the Gr\"uneisen parameter of mode $r, {\bf k}$ \cite{as76}. 

For $T \to 0$, Eqs.~(\ref{apt}) and (\ref{ert}) yield:
\begin{equation}
 A_p^{M}(0) - A_0   = \frac{1}{2 N B_0}
   \sum_{r, {\bf k}}  \hbar \omega_r({\bf k}) \gamma_r({\bf k})  \, .
\label{ap1}
\end{equation}
Taking into account that the frequencies $\omega_r({\bf k})$
scale with the mass as $M^{-1/2}$, Eq.~(\ref{ap1}) can be written as
\begin{equation}
   A_p^{M}(0) - A_0 = \frac{C}{M^{1/2}}    \, ,
\label{ap2}
\end{equation}
where $C$ is a constant independent of the mass.
Thus, the difference $\Delta A_p(0) = A_p^{M_2}(0) - A_p^{M_1}(0)$
for isotopic masses $M_2$ and $M_1$ can be written to first order:
\begin{equation}
 \Delta A_p(0) = \left. \frac{\partial A_p^M(0)} {\partial M} 
       \right|_{M_1}  \Delta M   =   
      - \frac12  \frac{C}{M_1^{3/2}}  \Delta M    \,  ,
\label{dap0}
\end{equation}
where $\Delta M  = M_2 - M_1$

At high temperature, Eq~(\ref{apt}) can be expanded by using
the expression (\ref{ekt}) in Appendix A for a harmonic oscillator.
This gives:
\begin{equation}
	A_p^M(T) = A_0 + C_1 N k_B T + \frac{C_2}{M} + O(M^{-2})   \, ,
\label{ap4}
\end{equation}
with $C_1$ and $C_2$ independent of the mass.
Then, the difference in area for two isotopes is
\begin{equation}
	\Delta A_p(T) = - \frac {C_2}{M_1^2}  \Delta M    \, .
\end{equation}
The classical limit ($M \to \infty$) is obtained
from Eq~(\ref{ap4}) as:
\begin{equation}
  A_p^{\rm cl}(T) = A_0 + C_1 N k_B T   \, .
\label{ap5}
\end{equation}

\section{Low-temperature isotopic effect in stressed bonds}

In general, isotopic effects take their maximum values for
$T \to 0$. For the C--C bond distance, we have at low temperature
from Eq.~(\ref{dd0}): 
\begin{equation}
  d_Q^M(0) = d_0 + \frac{Z}{d_0 \omega_{\rm eff} M_{\rm red} }   \, ,
\label{dm0_ap}
\end{equation}
with the stress-independent parameter
$Z = \hbar \gamma_{\rm eff} / 2$.
Taking into account that 
$\omega_{\rm eff} = (k_{\rm eff} / M_{\rm red})^{1/2}$,
it is clear that $d_Q^{12}(0) > d_Q^{13}(0)$, as shown in Fig.~5(a). 

In Eq.~(\ref{dm0_ap}), both the classical low-$T$ distance $d_0$ 
and the effective frequency $\omega_{\rm eff}$ change with
an applied in-plane stress $\tau$.
Thus, we have
\begin{equation}
 \frac {\partial d_Q^M(0)}{\partial \tau} = 
   \frac {\partial d_0}{\partial \tau} - 
   \frac{Z}{d_0 \omega_{\rm eff} M_{\rm red}}
   \left( \frac{1}{d_0} \frac {\partial d_0}{\partial \tau} + 
   \frac{1}{\omega_{\rm eff}} \frac {\partial \omega_{\rm eff}}
          {\partial \tau}  \right)   \, .
\label{partd_ap}
\end{equation}
It turns out that ${\partial d_0}/{\partial \tau} < 0$ and
$\partial \omega_{\rm eff} / \partial \tau > 0$, but the contribution
of the second term in the bracket of the r.h.s. in Eq.~(\ref{partd_ap}) 
is larger than that of the first one, so that the whole bracket
is positive and independent of the isotopic mass.

For not very large stress, and
within the effective-frequency approximation, the stress derivative
of $\omega_{\rm eff}$ can be written as
\begin{equation}
    \frac {\partial \omega_{\rm eff}} {\partial \tau} =
   \frac {\partial \omega_{\rm eff}} {\partial d_0}
   \frac {\partial d_0} {\partial \tau} = 
    - \gamma_{\rm eff} \frac {\omega_{\rm eff}} {d_0} 
   \frac {\partial d_0} {\partial \tau}   \, ,     
\label{partom_ap}
\end{equation}
where we have used the definition of $\gamma_{\rm eff}$ in
Eq.~(\ref{geff}).
Then, we have from Eq.~(\ref{partd_ap}):
\begin{equation}
  \frac {\partial d_Q^M(0)}{\partial \tau} =
   \frac {\partial d_0} {\partial \tau} -
   \frac {Z'}{d_0 \omega_{\rm eff} M_{\rm red}  \, ,}
\label{partd2_ap}
\end{equation}
with 
\begin{equation}
   Z' =  (1 - \gamma_{\rm eff})  
      \frac {\hbar \gamma_{\rm eff}}{2 d_0} 
      \frac {\partial d_0}{\partial \tau} \, .
\end{equation}
The parameter $Z'$ is positive (as $\gamma_{\rm eff} > 1$ and
$\partial d_0 /\partial \tau < 0$) and independent of 
the isotopic mass.

Putting $\omega_{\rm eff} = (k_{\rm eff} / M_{\rm red})^{1/2}$,
we obtain from Eq.~(\ref{partd2_ap}):
\begin{equation}
 \frac {\partial d_Q^{13}(0)}{\partial \tau} -
 \frac {\partial d_Q^{12}(0)}{\partial \tau} =   
     \frac {Z' A}{d_0}  > 0 \, ,
\label{partd3_ap}
\end{equation}
with
\begin{equation}
  A =  k_{\rm eff}^{-1/2}   \left[ (M_{\rm red}^{12})^{-1/2} - 
         (M_{\rm red}^{13})^{-1/2} \right] > 0   \, .
\label{aa_ap}
\end{equation}
This means that the difference $d_Q^{12}(0) - d_Q^{13}(0)$ is
reduced for increasing compressive stress and increases under
a tensile stress.

The low-temperature parameter $\Lambda_{\rm C-C}$ may be
written as
\begin{equation}
  \Lambda_{\rm C-C} = \frac {d_Q^{13}(0)}{d_Q^{12}(0)} - 1  \; .
\label{lambda2_ap}
\end{equation} 
Its change with an applied stress $\tau$ is given by the
derivative
\begin{equation}
  \frac{\partial \Lambda_{\rm C-C}}{\partial \tau} = 
          \frac{F_0}{d_Q^{12}(0)^2}  \, ,
\label{dlambda_ap}
\end{equation}
with
\begin{equation}
  F_0 = d_Q^{12}(0) \frac{\partial d_Q^{13}(0)}{\partial \tau} -
        d_Q^{13}(0) \frac{\partial d_Q^{12}(0)}{\partial \tau}   \; .
\end{equation}
It turns out that $F_0 > 0$, so 
$\partial \Lambda_{\rm C-C} / \partial \tau > 0$, and
$\Lambda_{\rm C-C}$ increases for a compressive stress.

This can be seen by using the expressions given above for
$d_Q^M(0)$ and $\partial d_Q^M(0) / \partial \tau$ in
Eqs.~(\ref{dm0_ap}) and (\ref{partd2_ap}). We obtain
\begin{equation}
  F_0 = A \, Z \left(  \frac{2}{d_0} \frac{\partial d_0}{\partial \tau} +
        \frac{1}{\omega_{\rm eff}} 
       \frac {\partial \omega_{\rm eff}} {\partial \tau} \right)  \, ,
\end{equation}
with $Z = \hbar \gamma_{\rm eff} / 2$ and $A$ defined
in Eq.~(\ref{aa_ap}), or
\begin{equation}
  F_0 =  (2 - \gamma_{\rm eff})   \frac{A \, Z}{d_0} \, 
         \frac{\partial d_0}{\partial \tau}  \, .
\end{equation}
Note that $F_0$ is positive for $\gamma_{\rm eff} > 2$ because
$\partial d_0 /\partial \tau < 0$, as happens
for C--C bonds in graphane. In the case 
$\gamma_{\rm eff} < 2$, one would have
$\partial \Lambda_{\rm C-C} / \partial \tau < 0$.

% ------------------------------------------------------------

%  BIBLIOGRAPHY

\end{document}